\documentclass[pre,multicol,aps,amsmath,amssymb]{revtex4}

\usepackage{graphicx}
\usepackage{dcolumn}
\usepackage{bm}

\usepackage{amssymb}
\usepackage{amsmath}
\begin{document}


\title{Soliton ratchets in homogeneous nonlinear Klein-Gordon
systems}

\author{Luis Morales--Molina}
\affiliation{Max-Planck Institut f\"ur Physik Komplexer Systeme,
N\"othnitzer Str. 38, 01187 Dresden, Germany}
\email{lmolina@mpipks-dresden.mpg.de}

\author{Niurka R.\ Quintero}
\affiliation{Departamento de F\'\i sica Aplicada I, E.U.P.,
Universidad
de Sevilla, c/ Virgen de \'Africa 7, 41011 Sevilla, Spain
}
\email{niurka@euler.us.es}

\author{Angel S\'anchez}%
\homepage{http://gisc.uc3m.es/~anxo}
\affiliation{%
Grupo Interdisciplinar de Sistemas Complejos (GISC) and
Departamento de Matem\'aticas, Universidad Carlos III de Madrid,
Avenida de la Universidad 30, 28911
Legan\'es, Madrid, Spain, and\\
Instituto de Biocomputaci\'on y F\'\i sica de Sistemas Complejos
(BIFI), Universidad de Zaragoza, 50009 Zaragoza, Spain}%

\author{Franz G.\ Mertens}
\email{Franz.Mertens@uni-bayreuth.de} \affiliation{Physikalisches
Institut, Universit\"at Bayreuth, D-85440 Bayreuth, Germany}

\date{\today}


\begin{abstract}
We study in detail the ratchet-like dynamics of topological
solitons in homogeneous nonlinear Klein-Gordon systems driven by a
bi-harmonic force. By using a collective coordinate approach with
two degrees of freedom, namely the center of the soliton, $X(t)$,
and its width, $l(t)$, we show, first, that energy is
inhomogeneously pumped into the system, generating as result a
directed motion; and, second, that the breaking of the time shift
symmetry gives rise to a resonance mechanism that takes place
whenever the width $l(t)$ oscillates with at least one frequency
of the external ac force. In addition, we show that for the
appearance of soliton ratchets, it is also neccesary to break the
time-reversal symmetry. We analyze in detail the effects of
dissipation in the system, calculating the average velocity of the
soliton as a function of the ac force and the damping. We find
current reversal phenomena depending on the parameter choice and
discuss the important role played by the phases of the ac force.
Our analytical calculations are confirmed by numerical simulations
of the full partial differential equations of the sine-Gordon and
$\phi^4$ systems, which are seen to exhibit the same qualitative
behavior. Our results show features similar to those obtained in
recent experimental work on dissipation induced symmetry breaking.
\end{abstract}

\maketitle

\noindent
{\bf LEAD PARAGRAPH}

\noindent Ratchet, or rectification, phenomena in nonlinear
nonequilibrium systems are receiving a great deal of attention in
the past few years. A typical example of a ratchet occurs when
point-like particles are driven by deterministic or non-white
stochastic forces: Under certain conditions related to the
breaking of symmetries, unidirectional motion can take place
although the applied force has zero average in time. The broken
symmetries can be spatial, by introducing an asymmetric potential,
or temporal, by using asymmetric periodic forces. This effect has
found many applications in the design of devices for rectification
or separation of different particles. Recently, ratchet dynamics
is being studied in the context of solitons, trying to understand
whether the fact that solitons, or nonlinear coherent excitations
in general, are extended objects, allows for similar rectification
phenomena. In nonlinear Klein-Gordon systems, it has been shown
that ratchet-like behavior can be observed when driven by
asymmetric periodic forces. However, the phenomenon is different
from that seen in point-like particles as the deformations of the
soliton play a crucial role. In this paper we describe in detail
these phenomena, analyze the mathematical conditions for its
existence, explain the underlying mechanism originating the net
motion, and study the highly non-trivial effects of dissipation on
the motion. Our results may be related to recent observations of
rectification in Josephson junctions and optical lattices.

\section{Introduction}

Ratchet or rectification phenomena, the subject of an intense
research effort during the last decade, appear in very many
different fields ranging from nanodevices to molecular biology
\cite{Hanggi1,Astumian,Ajdari,Reimann,Linke,Hanggi2,doering}.
Generally speaking, the appearance of ratchet-like behavior
requires two ingredients: Departure from thermal equilibrium,
either by using correlated stochastic forces or deterministic
forces, and breaking of spatial and/or temporal symmetries
\cite{Reimann,Hanggi2,asym,save,Borromeo,Marchesoni,maskk,Morales}.
Many applications of the ratchet theoretical framework deal with
the scenario of spatial symmetry breaking, beginning with the
proposal for nonequilibrium rectifiers in \cite{Hanggi1}. However,
in other problems, breaking the temporal symmetry may be much more
feasible; this is the case, for instance, of optical lattices
\cite{schia} or Josephson junctions \cite{Zapata,JJ1,JJ2} (see
also \cite{kk} for related work on Josephson ratchets in the
dissipative quantum regime) when driven by an appropriate
bi-harmonic force. In this context, the ratio of the frequencies
of a bi-harmonic force is related with the breaking of the
temporal-shift symmetry, (given by $f(t) = - f(t+T/2)$ with $T$
being the period) whereas suitable choices for the driving phases
and damping \cite{optical2} lead to time-reversal symmetry
breaking \cite{ricardo}. This mechanism to obtain directed motion
from a zero-mean force has been first proposed for particle
(zero-dimensional) systems \cite{Reimann,asym,referee-flach}, and
it has recently been extended to extended systems, both classical
\cite{flach,Salerno-mixing,Marchesoni,maskk} and quantum ones
\cite{Goychuk}.

A very relevant system, mostly because its character of paradigmatic of
phenomena arising in connection with topological solitons and its many
applications, is the damped and driven sine-Gordon (sG) model
\cite{Scott}:
\begin{equation}\label{sG}
\phi_{tt} - \phi_{xx} + \sin(\phi) = -\beta \phi_{t} + f(t).
\end{equation}
In \cite{Salerno-mixing} the symmetry considerations above were
studied in the context of Eq.\ (\ref{sG}) by choosing
\begin{equation}\label{ac}
f(t)=\epsilon_{1} \sin(\delta t + \delta_{0}) + \epsilon_{2}
\sin(m \delta  t + \delta_{0}+\theta') =\epsilon_{1} \sin(\delta t
+ \theta_{1}) + \epsilon_{2} \sin(m \delta  t + \theta_{2}),
\end{equation}
where $m$ is an integer number and $\epsilon_{1}$ ($\epsilon_{2}$)
is the amplitude of the harmonic component with frequency $\delta$
($m \delta$) and phase $\theta_{1}$ ($\theta_{2}$). We note that
in the equation above $\theta_{1}=\delta_{0}$ y
$\theta'=\theta_{2}-\theta_{1}$, being $\theta'$ the relative
phase, and therefore the results presented here can be understood
as well in terms of relative phases. The reason for including two
harmonics is that, as can be seen immediately from the previous
symmetry analysis, in the case of a single harmonic the directed
motion of sG kinks is not possible, something advanced by a
different reasoning some years ago \cite{niurka5} and
experimentally confirmed in \cite{goldo}. Thus, the main result in
\cite{Salerno-mixing} was the finding that, if $f(t)$ breaks the
shift symmetry {\em and if the total topological charge in the
system is nonzero}, a directed current (meaning a preferential
motion of the kink in one direction) could be observed whose
direction and magnitude depend on the driving parameters and
damping coefficient.

Subsequently, in a recent paper \cite{morales}, we advanced a
first explanation of the mechanisms underlying the observations in
\cite{Salerno-mixing}: Directed motion of solitons arises because
of a resonance phenomenon involving the oscillations of the
soliton width, $l(t)$, with at least one of the frequencies of the
ac force. In terms of the symmetry picture we are discussing here,
we note that it is precisely the breaking of the shift symmetry
the reason for the appearance of the resonance, which means that
the shift symmetry and the resonance conditions are equivalent. In
the present work we will extend our preliminary results in
\cite{morales} in several directions. First, we will show that net
motion of the soliton is only possible if, {\em in addition to the
aforementioned resonance condition on the frequencies}, the
time-reversal symmetry is also broken by the action of ac force
and damping. As will be seen below, this in turn implies that the
energy introduced by the biharmonic force is inhomogeneously
pumped into the system, giving rise to the observed unidirectional
motion. Second, we will carry out a detailed study of the ratchet
dynamics of solitons governed by Eq.\ (\ref{sG}), including not
only the cases $m=2$, our main focus in \cite{morales}, but also
$m=4$, which we compare to the previous one. Third, in order to
show the generality of our results, we will present an analysis of
the motion of the $\phi^4$ kink, whose dynamics is given by
\begin{equation}\label{phi4}
\phi_{tt} - \phi_{xx} + \phi^3 - \phi = -\beta \phi_{t} + f(t).
\end{equation}
Finally, we will extend our previous work to analyze in depth the
influence of dissipation in the directional motion. We will focus
on the way that damping affects the phase values for which there
is no soliton motion, and we will discuss the existence of values
of the dissipation coefficient that lead to optimal soliton
mobility. Interestingly, our results are reminiscent of the
experimental findings in \cite{optical2}, where similar effects
due to dissipation are found.

The presentation of the above results is organized as follows: In
the next Section we analyze twp nonlinear Klein-Gordon systems,
specifically the sine-Gordon and $\phi^4$ models, by means of a
collective coordinate (CC) approach, obtaining the analytical and
approximate expressions for the average of the velocity of the
center of the soliton for $m=2$  and $m=4$. In Section III we
verify our analytical results by means of numerical simulations.
Section IV is devoted to the discussion of the most relevant
effects of dissipation. To conclude the paper, in the last section
we recapitulate the results of Sections III--V, make the
connection with the experiments and summarize our main findings.

\section{Collective coordinate approach}  \label{CCA-section}

\subsection{Collective coordinate equations}

Our aim in this section is to obtain an approximate expression for
the average velocity of the center of the kink (as $\phi^4$ kinks
are not solitons, strictly speaking, we will refer to these
excitations of both models as kink from now on). To this end, we
will use one of the different CC approaches (see \cite{siam} for a
review). We introduce two variables, namely the center of the
kink, $X(t)$, and its width, $l(t)$, through Rice's {\it Ansatz}
\cite{rice}, which amounts to specifying the kink solution as
\begin{equation}\label{RicesG}
\phi_{0}(x,t) = 4\,{\rm arctan} \left[\exp
\left(\frac{x-X(t)}{l(t)} \right) \right],
\end{equation}
for sG and
\begin{equation}\label{Ricephi4}
\phi_{0}(x,t) = \tanh\left[\frac{x-X(t)}{l(t)}\right],
\end{equation}
for $\phi^4$. Using the above expressions, and
the variations of the energy and the
momentum (two conserved quantities of the unperturbed systems
Eqs.\ (\ref{sG}) and (\ref{phi4}))
it can be shown (see \cite{niurka2} for details) that
the two collective variables
obey the following two ordinary differential equations:
\begin{eqnarray} \label{cc-p}
& \, & \frac{dP}{dt}=-\beta P-qf(t), \\
& \, &  \dot{l}^2-2l\ddot{l}-2\beta l\dot{l}=
\Omega_{R}^2 l^2\left[1+\frac{P^2}{M_{0}^2}\right]-\frac{1}{\alpha},
\label{cc-l}
\end{eqnarray}
where
$P(t)\equiv M_{0} l_{0} \dot{X}/l(t)$,
$\Omega_{R} = 1/(\sqrt{\alpha} l_{0})$ is Rice's frequency, and
the parameters $M_{0}$, $q$, $\alpha$ and $l_{0}$ take different values
according to the model, sG or $\phi^4$ (see Table \ref{tab0}).
\begin{table}[h]
\begin{minipage}{0.6\textwidth}
\begin{tabular}{|c|c|c|}
\hline
 Parameters & sine-Gordon  & $\phi^4$ \\
\hline
$q$ & $2\pi$  & $2$ \\
\hline
$M_{0}$ & $8$ & $2 \sqrt{2}/3$\\
\hline
$l_{0}$ & $1$  &  $\sqrt{2}$\\
\hline
$\alpha$ & $\pi^{2}/12$ &  $(\pi^{2}-6)/12$\\
\hline
\end{tabular}
\end{minipage}
\hspace{0.6cm} \caption{Parameters corresponding to the two
nonlinear Klein-Gordon models we work with, the sine-Gordon and
$\phi^4$ models. $q$ is the topological charge, $M_{0}$ and
$l_{0}$ are the mass and the width of the kink at
rest.}\label{tab0}
\end{table}

The quantity $P(t)$ deserves some discussion on in its own. To
begin with, it represents the momentum of the kink: Indeed, the
expression for $P(t)$ can be obtained as well by substituting
Rice's {\it Ansatz} into the definition of the momentum
$P(t)=-\textstyle \int_{-\infty}^{+\infty} dx \,\ \phi_{t}
\phi_{x}$. The same equations can be obtained using a projection
technique with a {\em Generalized Traveling Wave Ansatz} (GTWA)
(see details in \cite{niurka3}). In addition, it has been shown
recently \cite{comment} that Eq.\ (\ref{cc-p}) for the momentum,
obtained here within the framework of the Rice approximation, {\it
is an exact equation of motion}, irrespective of the choice for
the {\it Ansatz}. Hence, its solution is also exact and not an
approximation.

Let us now proceed with the analysis of Eqs.\ (\ref{cc-p}) and
(\ref{cc-l}).
After transients have elapsed and become negligible,
($t\gg 1/\beta$) the solution for $P(t)$ is essentially given by
\begin{equation}\label{2.6}
\ P(t)=-\sqrt{\epsilon} [a_{1}\sin(\delta t+\theta_{1}-\chi_{1})+
a_{2}\sin(m\delta t+\theta_{2}-\chi_{m})],
\end{equation}
where $\epsilon\equiv\displaystyle \mbox{min}(\epsilon_1,\epsilon_2)$
is a rescaling parameter, and
\begin{eqnarray}
\displaystyle \chi_{1}&=&\arctan\left(\frac{\delta}{\beta}\right),
\quad \chi_{m}=\arctan\left(\frac{m\delta}{\beta}\right), \\
\displaystyle a_{1}&=&\frac{q}{\sqrt{\beta^2+\delta^2}}
\frac{\epsilon_{1}}{\sqrt{\epsilon}},\quad{\rm and}\quad
\displaystyle a_{2}=
\frac{q}{\sqrt{\beta^2+m^2\delta^2}}\frac{\epsilon_{2}}{\sqrt{\epsilon}}.
\end{eqnarray}
We can now turn to Eq.\ (\ref{cc-l}), which is nothing but a
parametrically driven oscillator: it is clear that $l(t)$ is
indirectly driven by the harmonic forces with frequencies $\delta$
and $m\, \delta$, due to the term  $P^2$ on the r.h.s of Eq.\
(\ref{cc-l}). To solve (approximately) this equation, we expand
$l(t)$ in powers of $\epsilon$ around the unperturbed kink width
$l_{0}$:
\begin{equation}\label{2.8}
l(t)=l_{0}+\epsilon l_{1}(t)+\epsilon^2 l_{2}(t)+...\,\,\ .
\end{equation}
Substituting Eq.\ (\ref{2.8}) into Eq.\ (\ref{cc-l}) we find a
hierarchy of equations for different orders of powers in
$\epsilon$. The first three equations for $l_i(t)$ are given by:
\begin{eqnarray}\label{2.9}
\ddot{l}_{1}(t)+\beta \dot{l}_{1}(t)+\Omega_{R}^2 l_{1}(t)
&=&-\frac{\Omega_{R}^2}{2\epsilon M_{0}^{2}}P^{2}(t) l_{0}, \\
\label{2.10}
\ddot{l}_{2}(t)+\beta \dot{l}_{2}(t)+\Omega_{R}^2 l_{2}(t)
&=&-\frac{\Omega_{R}^2}{2\epsilon M_{0}^{2}}P^{2}(t) l_{1}+
\frac{\dot{l}_{1}^{2}}{2 l_{0}}+\frac{\Omega_{R}^2 l_{1}^{2}}{2 l_{0}}\\
\label{2.11}
\ddot{l}_{3}(t)+\beta \dot{l}_{3}(t)+\Omega_{R}^2 l_{3}(t)&=&
-\frac{\Omega_{R}^2}{2\epsilon M_{0}^{2}}P^{2}(t) l_{2}+
\frac{\dot{l}_{1}\dot{l}_{2}}{l_{0}}+\frac{\Omega_{R}^2 l_{1} l_{2}}
{l_{0}}-\frac{\dot{l}_{1}^{2} l_{1}}{2 l_{0}^2}-
\frac{\Omega_{R}^2 l_{1}^3}{2 l_{0}^2}.
\end{eqnarray}
These equations are linear and therefore we can begin by
solving the first one, Eq.\ (\ref{2.9}),
by substituting the expression for the
momentum, Eq.\ (\ref{2.6}), into the r.h.s of (\ref{2.9}).
With this, the equation for $l_{1}$ becomes
\begin{eqnarray}\label{2.12}
\nonumber \ddot{l}_{1}(t)+\beta \dot{l}_{1}(t)\Omega_{R}^2
l_{1}(t)&=& A_{1}+A_{2}\cos(2\delta t+2\theta_{1}-2\chi_{1})
\nonumber +A_{3}\cos(2m\delta t+2\theta_{2}-2\chi_{m})+\\&+&
A_{4}\Big\{\cos[(m-1)\delta t +\theta_{2}-\theta_{1}
-(\chi_{m}-\chi_{1})] -\cos[(m+1)\delta t+ \theta_{1}+ \theta_{2}
-(\chi_{m}+\chi_{1})]\Big\},
\end{eqnarray}
where
\begin{eqnarray}
A_{1}&=&-A_{2}-A_{3}, \\
A_{2}&=&\frac{\Omega_{R}a_{1}^2}{4\sqrt{\alpha }M_{0}^2}, \\
A_{3}&=&\frac{\Omega_{R}a_{2}^2}{4\sqrt{\alpha }M_{0}^2}, \\
A_{4}&=&-\frac{\Omega_{R}}{2\sqrt{\alpha }M_{0}^2}a_{1}a_{2}.
\end{eqnarray}

It is important to notice on the r.h.s of Eq.\ (\ref{2.12}) the
presence of harmonics with frequencies $2 \delta$, $2 m \delta$
and $(m \pm 1) \delta$. As a consequence, all these frequencies
appear in the expression for $l_{1}(t)$ and they remain after a
transient time $t \gg 1/\beta$, i.e., asymptotically we have
\begin{eqnarray} \nonumber
l_{1}(t)=&&\frac{A_{1}}{\Omega_{R}^2}+
\frac{A_{2}\sin(2\delta t+2\theta_{1}-2\chi_{1}+\tilde{\theta}_{2})}
{\sqrt{(\Omega_{R}^2-4\delta^2)^2+4\beta^2\delta^2}}+
\frac{A_{3}\sin(2m\delta t+2\theta_{2}-2\chi_{m}+
\tilde{\theta}_{2m})}{\sqrt{(\Omega_{R}^2-4m^2\delta^2)^2+
4m^2\beta^2\delta^2}}\\ \nonumber
&&+\frac{A_{4}\sin[(m-1)\delta t +\theta_{2}-\theta_{1} -(\chi_{m}-\chi_{1})+
\tilde{\theta}_{m-1}]}{\sqrt{(\Omega_{R}^2-(m-1)^2\delta^2)^2+
\beta^2(m-1)^2\delta^2}}
\\
&&-\frac{A_{4}\sin[(m+1)\delta t + \theta_{1}+
\theta_{2} -(\chi_{m}+\chi_{1})+\tilde{\theta}_{m+1}]}
{\sqrt{(\Omega_{R}^2-(m+1)^2\delta^2)^2+\beta^2(m+1)^2\delta^2}}, \label{2.13}
\end{eqnarray}
where
$$\tilde{\theta}_{m}=\arctan\left(\frac{\Omega_{R}^2-m^2\delta^2}{m\beta\delta}\right).$$
From the r.h.s of Eqs.\ (\ref{2.10}) and
(\ref{2.11}), the same reasoning as above leads one to
find the harmonics corresponding to the second and third
order corrections $l_2(t)$ and $l_3(t)$. We do not include the explicit
expressions for these corrections for the sake of brevity, but in Table
\ref{tab1} we collect all these values for $m=2, 3, 4$.
\begin{table}[h]
\begin{center}
\begin{tabular}{|c|c|c|}
\hline
 $2^{nd}$ harmonic($\delta$) & $l_{1}$ & $l_{2}$ \\
\hline
m & $2\delta$, $2m\delta$, $(m\pm 1)\delta$ & $2\delta$, $4\delta$, $4m\delta$, $(m\pm 1)\delta$, \\
\, & \, & $2(m\pm 1)\delta$, $(m\pm 3)\delta$, $(3m\pm 1)\delta$\\
\hline
2 & $\delta$, $2\delta$, $3\delta$, $4\delta$ & $\delta$, $2\delta$, $3\delta$, $4\delta$, $5\delta$, $6\delta$, $7\delta$, $8\delta$\\
\hline
3 & $2\delta$, $4\delta$, $6\delta$  &  $2\delta$, $4\delta$, $6\delta$, $8\delta$, $10\delta$,  $12\delta$ \\
\hline
4 & $2\delta$, $3\delta$, $5\delta$, $8\delta$ &  $\delta$,  $2\delta$, $3\delta$,  $4\delta$, $5\delta$, $7\delta$,\\
\, & \, & $9\delta$, $10\delta$, $11\delta$, $13\delta$, $16\delta$\\
\hline
\end{tabular}
\end{center}
\caption{Harmonic content of the first contributions to the
perturbative expansion of $l(t)$. Notice that $\delta$ and $m \delta$
are the frequencies of the ac force (they appear in the momentum
equation as well).
}\label{tab1}
\end{table}

\subsection{Velocity of the kink center}

Having found the exact expression for $P(t)$ and the first terms
in the perturbative expansion for $l(t)$, we can now calculate the
average velocity over one period $T=2 \pi/\delta$. To this end, we
recall the previously defined expression for the momentum,
$P(t)=M_{0} l_{0} \dot{X}/l(t)$. After transients have elapsed,
which means after a time essentially given by the inverse of the
dissipation coefficient, the mean velocity of the kink can be
expressed in terms of the CC as
\begin{equation}\label{2.14}
\langle\dot{X}(t)\rangle=\frac{1}{T}\int_{0}^{T}\frac{P(t)l(t)}{M_{0}l_{0}}dt.
\end{equation}
Taking into account the expansion (\ref{2.8}), this expression
can be written as
\begin{eqnarray}\label{2.15}
\nonumber
\langle\dot{X}(t)\rangle&=&\frac{1}{T}\int_{0}^{T}\frac{P(t)(l_{0}+
\epsilon l_{1}(t)+\epsilon^2 l_{2}(t)+...)}{M_{0}l_{0}}dt\\
&=&\langle\dot{X}_{0}(t)\rangle+\epsilon \langle\dot{X}_{1}(t)\rangle+\epsilon^2 \langle\dot{X}_{2}(t)\rangle+...
\end{eqnarray}
Therefore, as already stated, the mean velocity can be
analytically (and approximately) calculated taking into account
the exact solution for the momentum (\ref{2.6}) and the first
terms of the expansion for the width of the kink. At order
$O(\epsilon^{0})$, $\langle\dot{X}_{0}(t)\rangle \sim \langle P(t)
\rangle=0$ since the average of the momentum is zero [see Eq.\
(\ref{2.6})]. Accordingly, the net motion of the kink can only
arise from the contribution of higher order terms; hence, we
proceed by computing the integral defining
$\epsilon\langle\dot{X}_{1}(t)\rangle$. Inserting Eqs.\
(\ref{2.6}) and (\ref{2.13}) in (\ref{2.15}), straightforward
calculations show for $m=2$ that
\begin{eqnarray}\label{2.16}
\epsilon\langle\dot{X}_{1}\rangle&=&
\frac{q^3\Omega_{R}^2\epsilon_{1}^2\epsilon_{2}}{8M_{0}^{3}
(\beta^2+\delta^2)\sqrt{\beta^2+4\delta^2}}\left(\frac{2\cos[2\theta_{1}-
\theta_{2}+(\chi_{2}-2\chi_{1})-\tilde{\theta}_{1}]}{\sqrt{(\Omega_{R}^2-
\delta^2)^2+\beta^2\delta^2}} \right. \nonumber
\\
&&\left.
-\frac{\cos[2\theta_{1}-\theta_{2}+(\chi_{2}-2\chi_{1})+
\tilde{\theta}_{2}]}{\sqrt{(\Omega_{R}^2-4\delta^2)^2+
4\beta^2\delta^2}}\right).
\end{eqnarray}
It is important to recall that for the CC approach to be valid, we
must require that the kink shape is not very much distorted, and
hence that $\epsilon_{i}\ll 1$ and $\delta \ll 1$. Moreover, we
would like to stress that expression (\ref{2.16}) is valid only
for any finite but nonzero value of $\beta$, since we have
obtained it for $t>>1/\beta$.

We can recast Eq.\ (\ref{2.16}) in a simpler form to make more
transparent the dependencies on the different parameters, as
follows:
\begin{eqnarray}\label{2.16a}
\epsilon\langle\dot{X}_{1}\rangle&=&
\epsilon_{1}^{2} \, \epsilon_{2} \,
\Gamma_{1}(\beta, \delta) \sin[\theta_{2}
-2\theta_{1}+
 \Phi_{1}(\beta, \delta)],
\end{eqnarray}
where we have introduced an amplitude, $\Gamma_{1}$, whose
specific form is not needed and is therefore omitted for brevity,
and a phase,
\begin{equation}
\Phi_{1}(\beta,\delta)=
2\chi_1-\chi_2 -\arctan\left[\frac{2 \beta \delta
    [(\Omega_{R}^2-4\delta^2)^2+3\beta^2\delta^2-
    (\Omega_{R}^2-\delta^2)^2]}{2(\Omega_{R}^2-\delta^2)
    [(\Omega_{R}^2-4\delta^2)^2+4\beta^2\delta^2]+(\Omega_{R}^2-4\delta^2)
    [(\Omega_{R}^2-\delta^2)^2+\beta^2 \delta^2]}\right]
    \label{fase-damp},
\end{equation}
which are only functions of the dissipation coefficient and the
frequencies of the ac force. We note here that a similar
expression, albeit starting from a less general {\it Ansatz}, has
been obtained in \cite{Salerno-mixing}. In this new form, it is
apparent that the averaged velocity is a sinusoidal function of
$\theta_{1}$ and $\theta_{2}$, and it is proporcional to
$\epsilon_{1}^2\, \epsilon_{2}$. It is interesting to note that
the same dependence on amplitudes and phases is also found in
other systems, in principle unrelated with the ratchet dynamics of
solitons we are discussing here (see \cite{ricardo} and references
therein). On the other hand, from the experimental viewpoint, it
is worth noting that, in the limit $\delta \ll \beta$, we find
$\langle V \rangle \sim \sin(2\theta_{1}-\theta_{2})$, in
agreement with the experiments on JJ reported in \cite{JJ2}.
\begin{figure}
\begin{center}
\begin{tabular}{cc}
\hspace{-1cm}
\includegraphics[width=9.cm,height=8cm]{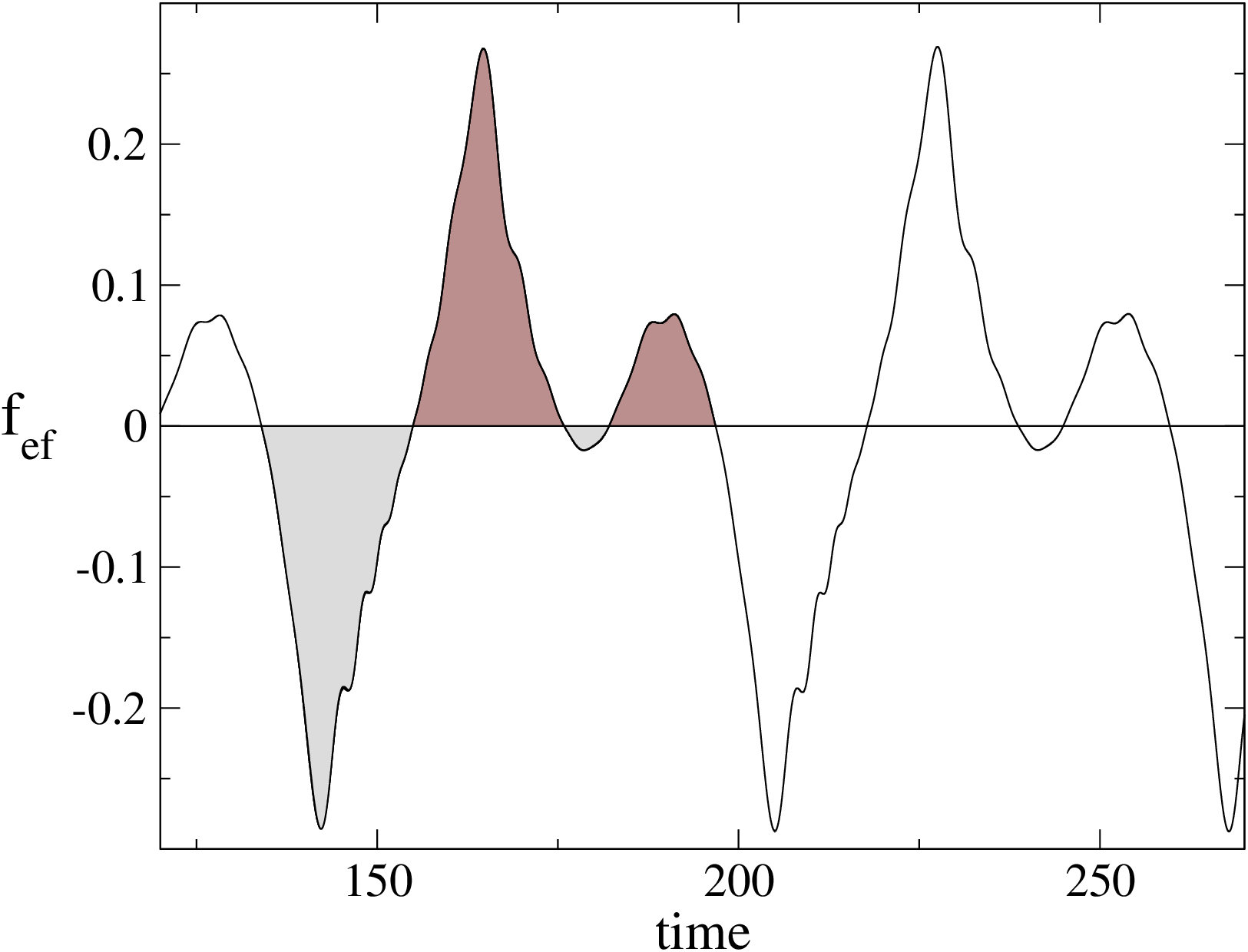}
\includegraphics[width=9.cm,height=8cm]{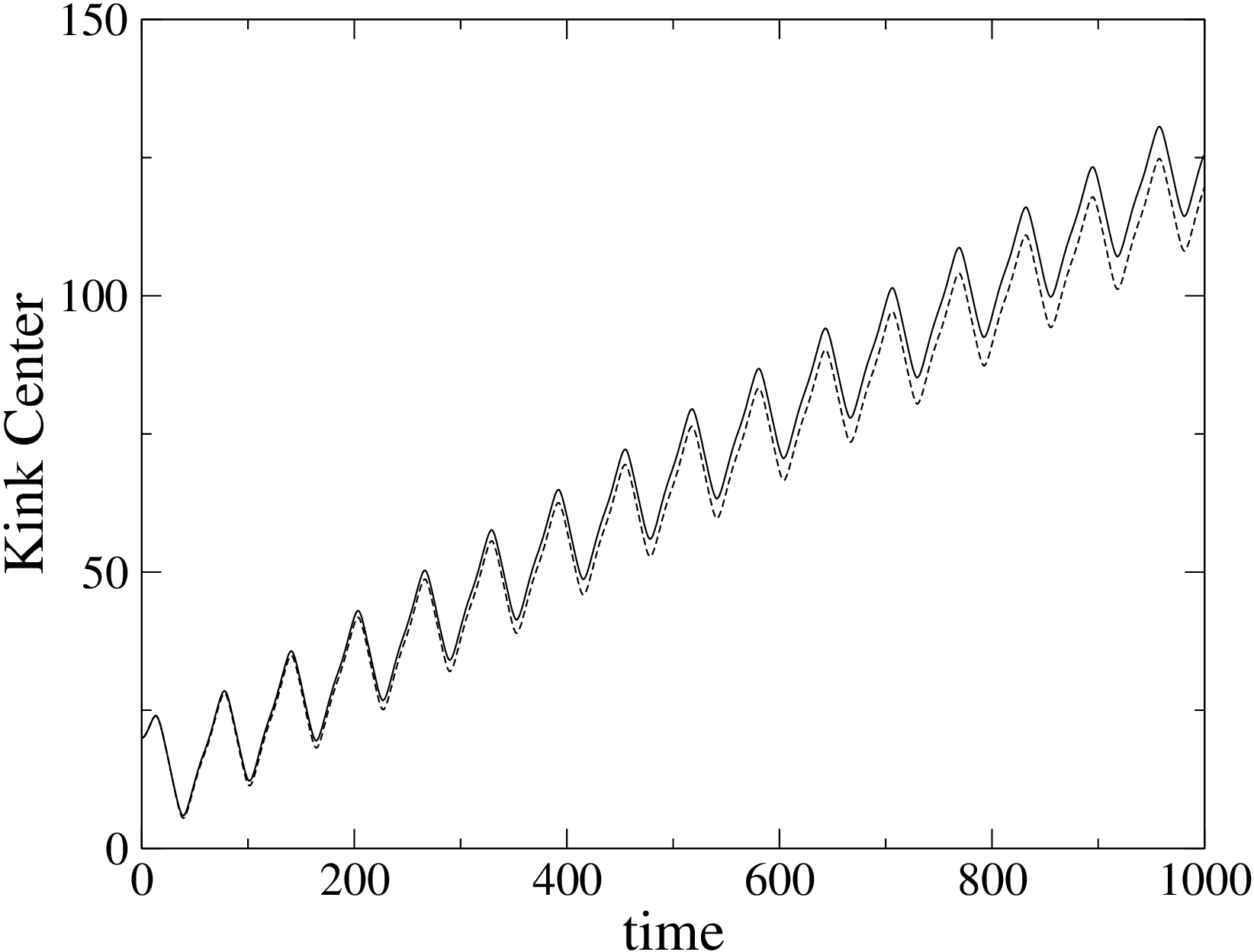}
\end{tabular}
\end{center}
\caption{a) Effective driving force in the sG model as a function
of time. b) Trajectory of the center of the soliton; dashed line:
CC result obtained
  by numerically integrating Eqs.\ (\ref{cc-p}) and (\ref{cc-l}); solid
  line: simulations of the full sG equation.
Parameters used in both figures are $\delta=0.1$, $m=2$,
$\beta=0.05$, $\theta_{2}=-2.5$,
  $\theta_{1}=0$, $\epsilon_{1}=\epsilon_{2}=0.2.$}
\label{effec-force}
\end{figure}
\begin{figure}
\begin{center}
\begin{tabular}{cc}
\hspace{-1cm}
\includegraphics[width=7.5cm,height=7cm]{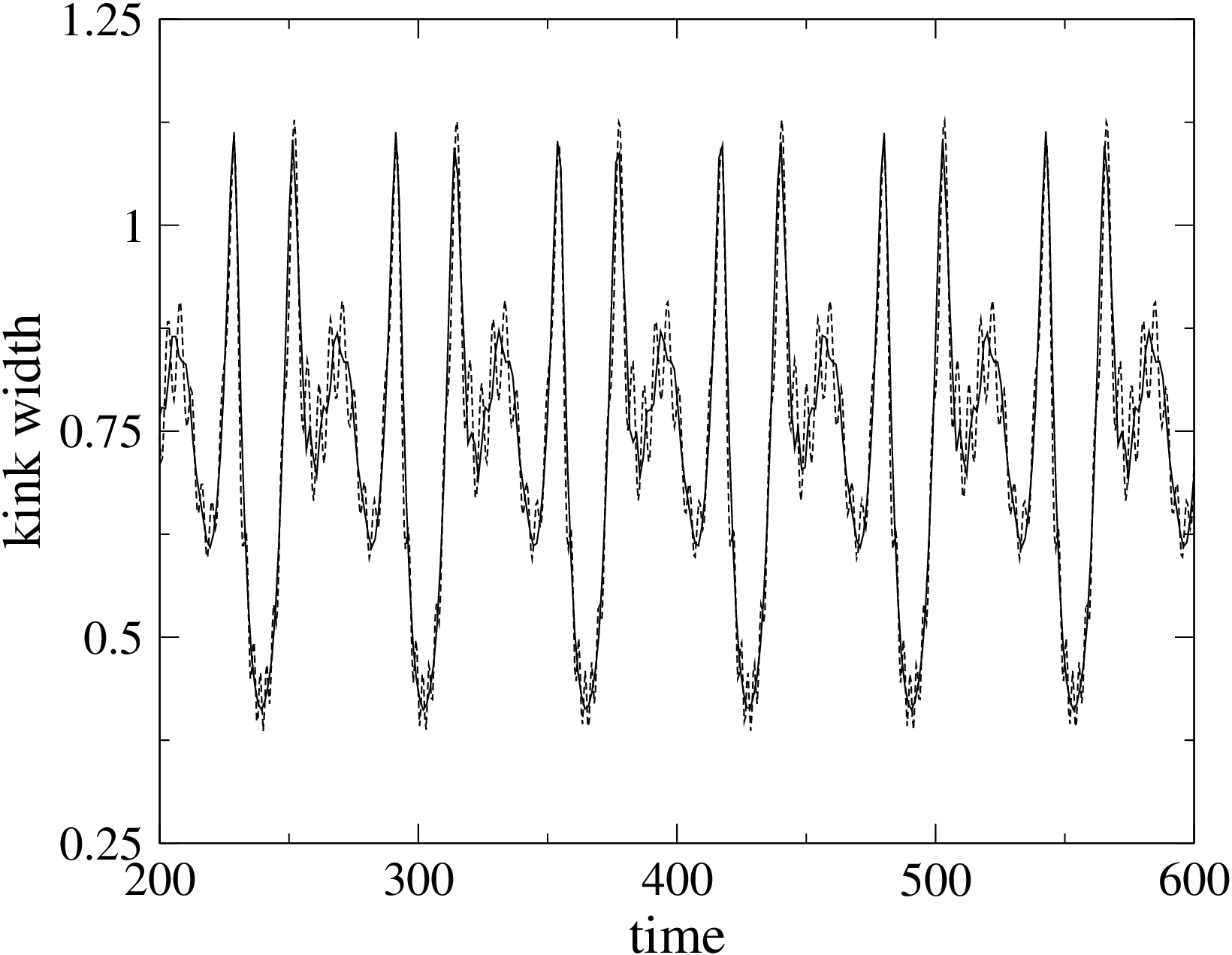}
\includegraphics[width=7.5cm,height=7cm]{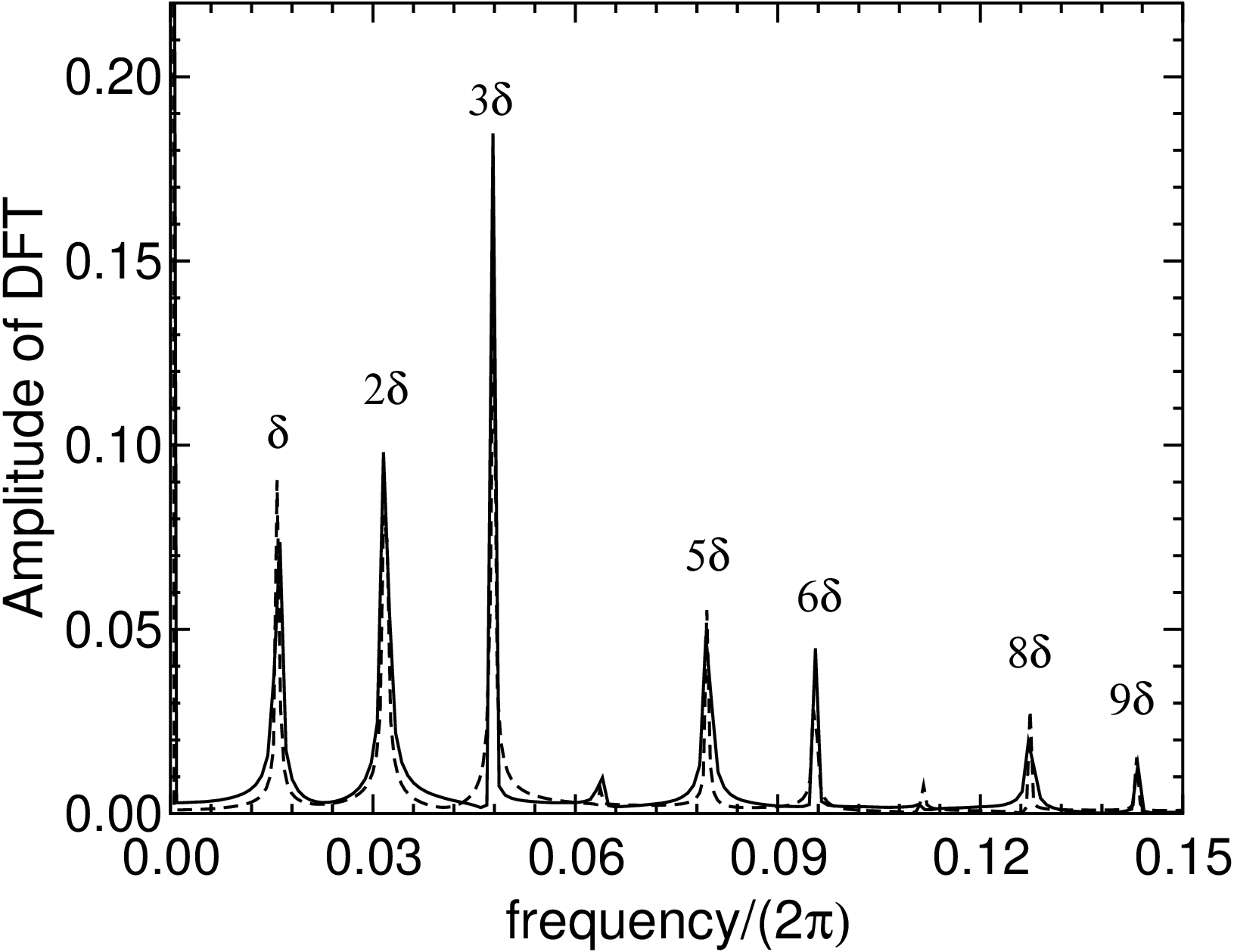}
\end{tabular}
\end{center}
\caption{Ratchet dynamics in the sG model for $m=2$. Left panel:
Width of the kink vs time. Right panel: Discrete Fourier Transform
(DFT) of the width of the kink. In both panels the dashed lines
represent the
 numerical solution of  Eqs.\ (\ref{cc-p}-\ref{cc-l}) and the solid
 lines the computed width from the numerical simulation of Eq.\ (\ref{sG}).
Parameters are $\epsilon_1=\epsilon_2=0.2$, $\beta=0.05$,
$\delta=0.1$, $\theta_{1}=-2.5$ $\theta_{2}=\pi/2-2.5$.}
\label{width-mixing}
\end{figure}
\begin{figure}
\begin{center}
\begin{tabular}{cc}
\hspace{-1cm}
\includegraphics[width=7.5cm,height=7cm]{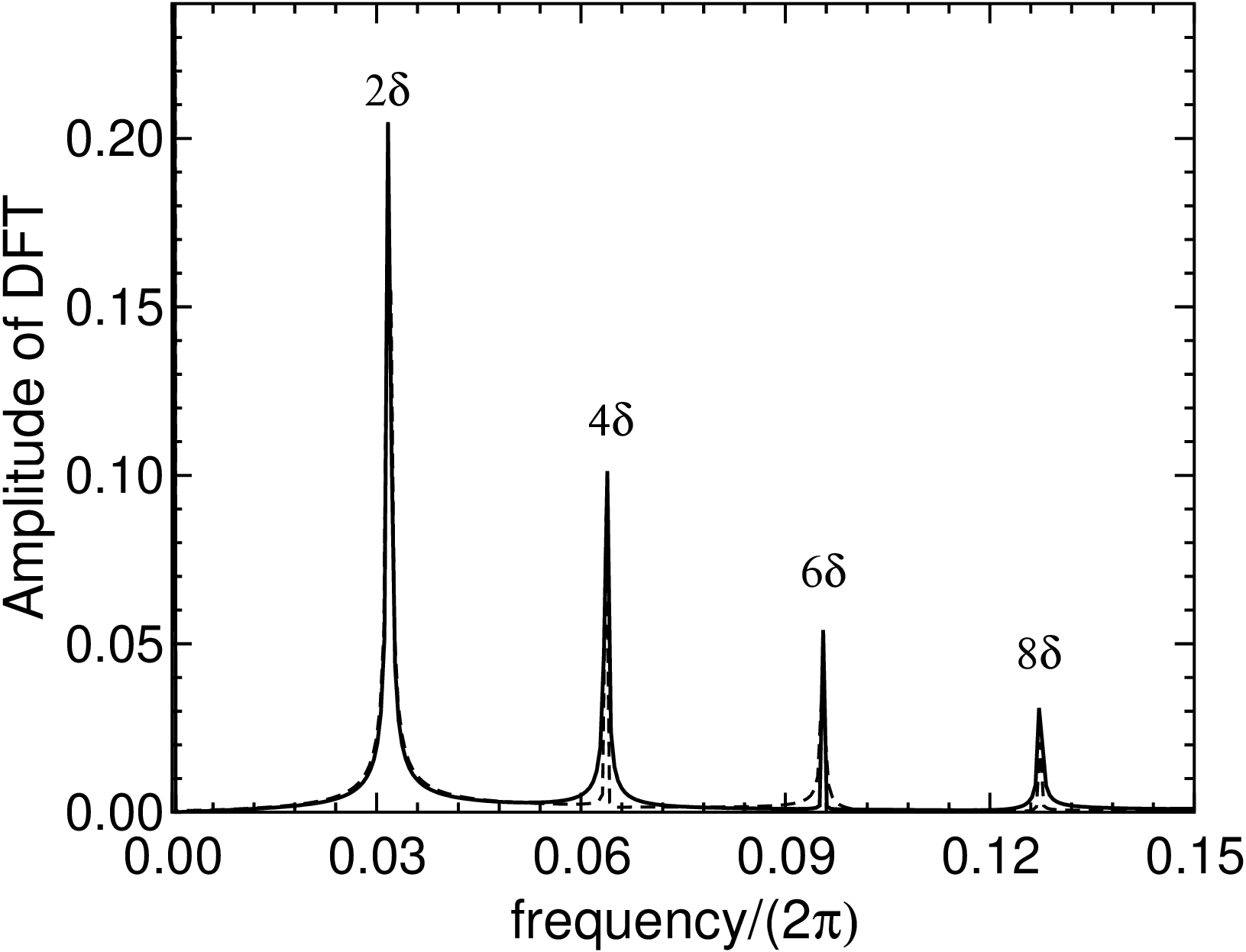}
\includegraphics[width=7.5cm,height=7cm]{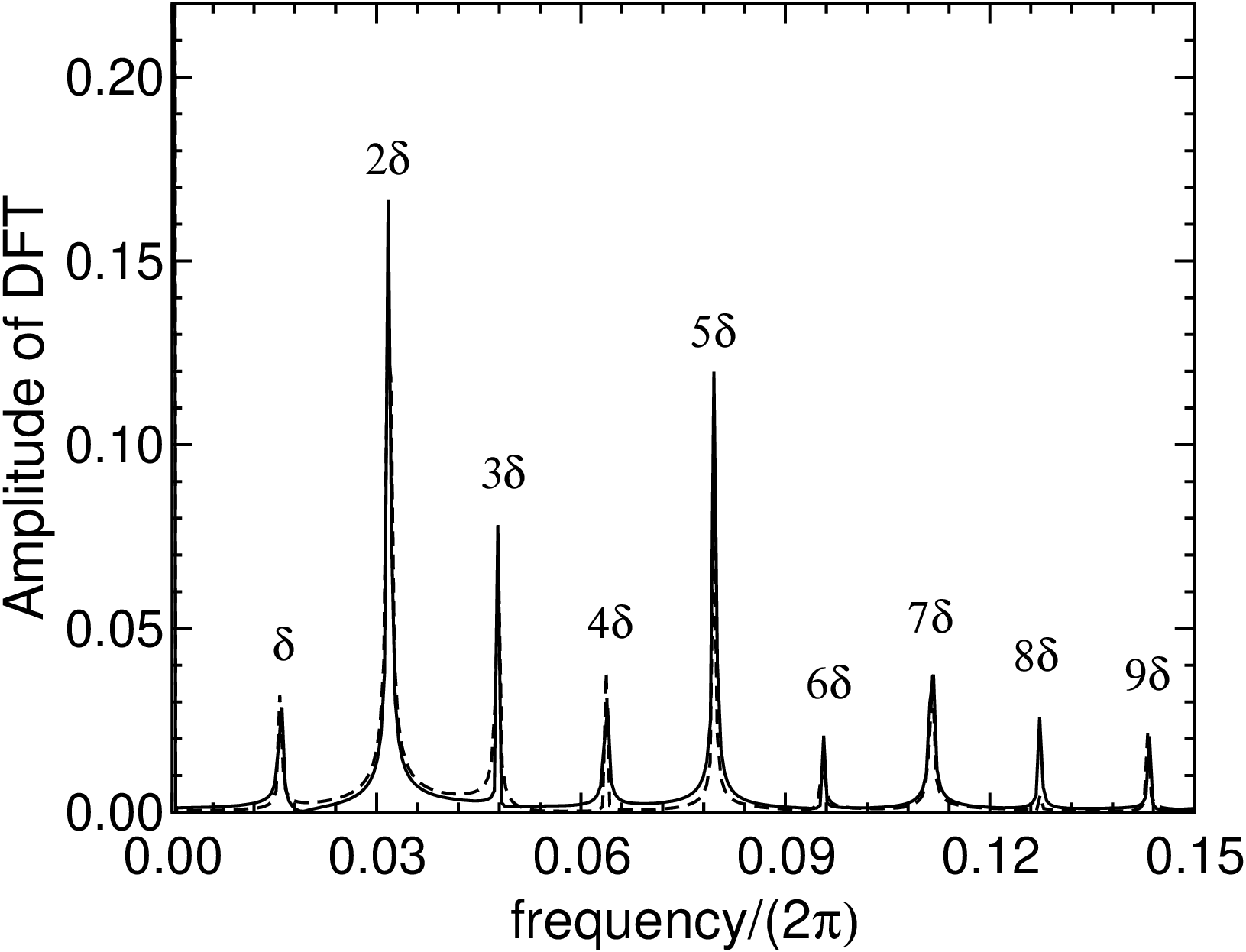}
\end{tabular}
\end{center}
\caption{\label{fig:dftsG} Discrete Fourier Transform of the kink
width. Left panel: $m=3$ (no net motion); Right panel: $m=4$
(ratchet motion of the kink takes place). Solid line: amplitude
measured in simulations. Dashed line: amplitude as obtained by
numerical integration of the CC equations
(\ref{cc-p})-(\ref{cc-l}). The parameters of the ac force and the
dissipation coefficient are the same as in Fig.
\ref{width-mixing}.}
\end{figure}

Proceeding with the analysis of the average velocity of the kink,
from Eq.\ (\ref{2.16a}) it is obvious that
net motion (i.e., with nonzero average velocity) is only possible if
\begin{equation}\label{src}
2 \theta_{1} - \theta_{2} -\Phi_{1} \ne n \pi, n=0, 1, ...
\end{equation}
In this form, this condition, as will be shown below (and was
pointed out in \cite{optical2}), arises from the breaking of the
time-reversal symmetry by the biharmonic force. Indeed, $\Phi_{1}$
is related with the difference of phases between the harmonics
with frequencies $\delta$ and $m \delta$ in the functions $P(t)$
and $l_{1}(t)$. Thus, our first resonance criterion reads that the
ratchet effect appears when $l(t)$ oscillates with at least one of
the frequencies of $P(t)$; in addition, the difference of phases
between the harmonics must be appropriate for the appeareance of
net motion. In others words, fixing all the parameters of the ac
force and damping except one of the two phases of the ac force,
one can find a critical value of that phase that suppresses
directed motion of kinks. The values at which the velocity
vanishes are important, since for a force with a given frequency
and damping, they indicate the phases at which the velocity of the
kink is reversed. On the other hand, the changes in the phases are
related with the breaking of the time-reversal symmetry of the
system \cite{denisov}.

Let us now turn to the case $m=4$, a value for which the
appearance of the ratchet effect is also predicted
\cite{Salerno-mixing}. In analogous manner to the preceding
calculations, one can show that
$\epsilon\langle\dot{X}_{1}\rangle$ is zero. Therefore, the
frequencies in $l(t)$ that contribute to the net motion must
appear in higher order corrections (see Table \ref{tab1}), namely
$\epsilon^2 \langle\dot{X}_{2}\rangle$. After cumbersome
calculations we find
\begin{eqnarray}\label{2.17}
\epsilon^2\langle\dot{X}_{2}\rangle&=&\frac{q^5\Omega_{R}^4\epsilon_{1}^4 \epsilon_{2}}{32M_{0}^5(\beta^2+\delta^2)^2 \sqrt(\beta^2+16\delta^2)}\,\ \mbox{x}\nonumber \\
&&\left\{\frac{\sin(4\theta_{1}-\theta_{2}+\chi_{4}-4\chi_{1}+\tilde{\theta}_{2}-
\tilde{\theta}_{1})}{\sqrt{(\Omega_{R}^2-\delta^2)^2+\beta^2\delta^2} %
\sqrt{(\Omega_{R}^2-4\delta^2)^2+4\beta^2\delta^2}} \right.
\nonumber \\
&&\left. +\frac{(6\delta^2+\Omega_{R}^2)\cos(4\theta_{1}-\theta_{2}+\chi_{4}-4\chi_{1}-\tilde{\theta}_{3}+\tilde{\theta}_{2}-%
\tilde{\theta}_{1})}{\sqrt{(\Omega_{R}^2-\delta^2)^2+\beta^2\delta^2} %
\sqrt{(\Omega_{R}^2-4\delta^2)^2+4\beta^2\delta^2}\sqrt{(\Omega_{R}^2-9\delta^2)^2+9\beta^2\delta^2}}\right. \nonumber
\\ %
&&\left. - \frac{\sin(4\theta_{1}-\theta_{2}+\chi_{4}-4\chi_{1}-\tilde{\theta}_{3}-\tilde{\theta}_{1}%
)}{\sqrt{(\Omega_{R}^2-\delta^2)^2+\beta^2\delta^2} %
\sqrt{(\Omega_{R}^2-9\delta^2)^2+9\beta^2\delta^2}}\right. \nonumber
\\ %
&&\left. -\frac{\sin(4\theta_{1}-\theta_{2}+\chi_{4}-4\chi_{1}+\tilde{\theta}_{2}+\tilde{\theta}_{4})}%
{2\sqrt{(\Omega_{R}^2-4\delta^2)^2+4\beta^2\delta^2} %
\sqrt{(\Omega_{R}^2-16\delta^2)^2+16\beta^2\delta^2}}\right. \nonumber
\\ %
&&  -\frac{(4\delta^2-\Omega_{R}^2)\cos(4\theta_{1}-\theta_{2}+\chi_{4}-4\chi_{1}+2\tilde{\theta}_{2}+\tilde{\theta}_{4})}%
{4[(\Omega_{R}^2-4\delta^2)^2+4\beta^2\delta^2] %
\sqrt{(\Omega_{R}^2-16\delta^2)^2+16\beta^2\delta^2}}
\Bigg \},
\end{eqnarray}
which can be rewritten as
\begin{eqnarray}\label{2.17a}
\epsilon^2\langle\dot{X}_{2}\rangle&=& \epsilon_{1}^4 \epsilon_{2}
\Gamma_{2}(\beta, \delta) \sin[\theta_{2}
-4\theta_{1}+
 \Phi_{2}(\beta, \delta)],
\end{eqnarray}
where again the amplitude, $\Gamma_{2}$, and the phase, $\Phi_{2}$,
are functions of the dissipation and the frequencies of the ac force.
As for the mobility condition, which for the $m=2$ case was given by
Eq.\ (\ref{src}), a similar expression
can be derived for the case $m=4$ by imposing that the sine function
vanishes in Eq.\ (\ref{2.17a}).
\begin{figure}
\begin{center}
\begin{tabular}{cc}
\includegraphics[width=6.5cm,height=7cm]{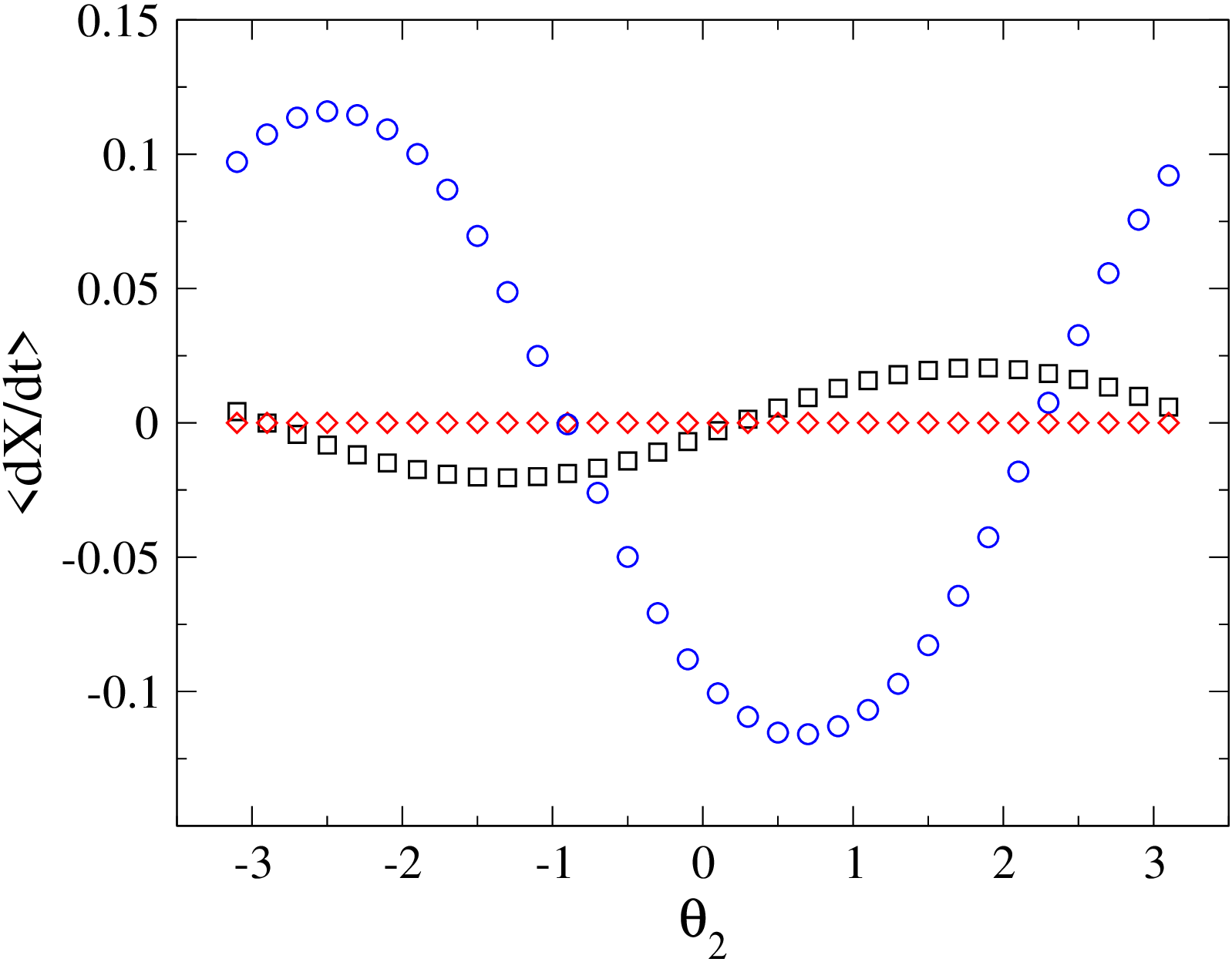}
 \end{tabular}
\end{center}
\caption{Dependence of the average velocity of a sG kink on the
phase, $\theta_{2}$, for different values of $m$. $m=2$ (blue
circles); $m=3$ (red diamonds) and $m=4$ (black squares).
Parameters are $\epsilon_1=\epsilon_2=0.2$, $\beta=0.05$,
$\delta=0.1$, $\theta_{1}=0$.} \label{sGm2m3m4}
\end{figure}
\begin{figure}
\begin{center}
\begin{tabular}{cc}
\includegraphics[width=7.5cm,height=7cm]
{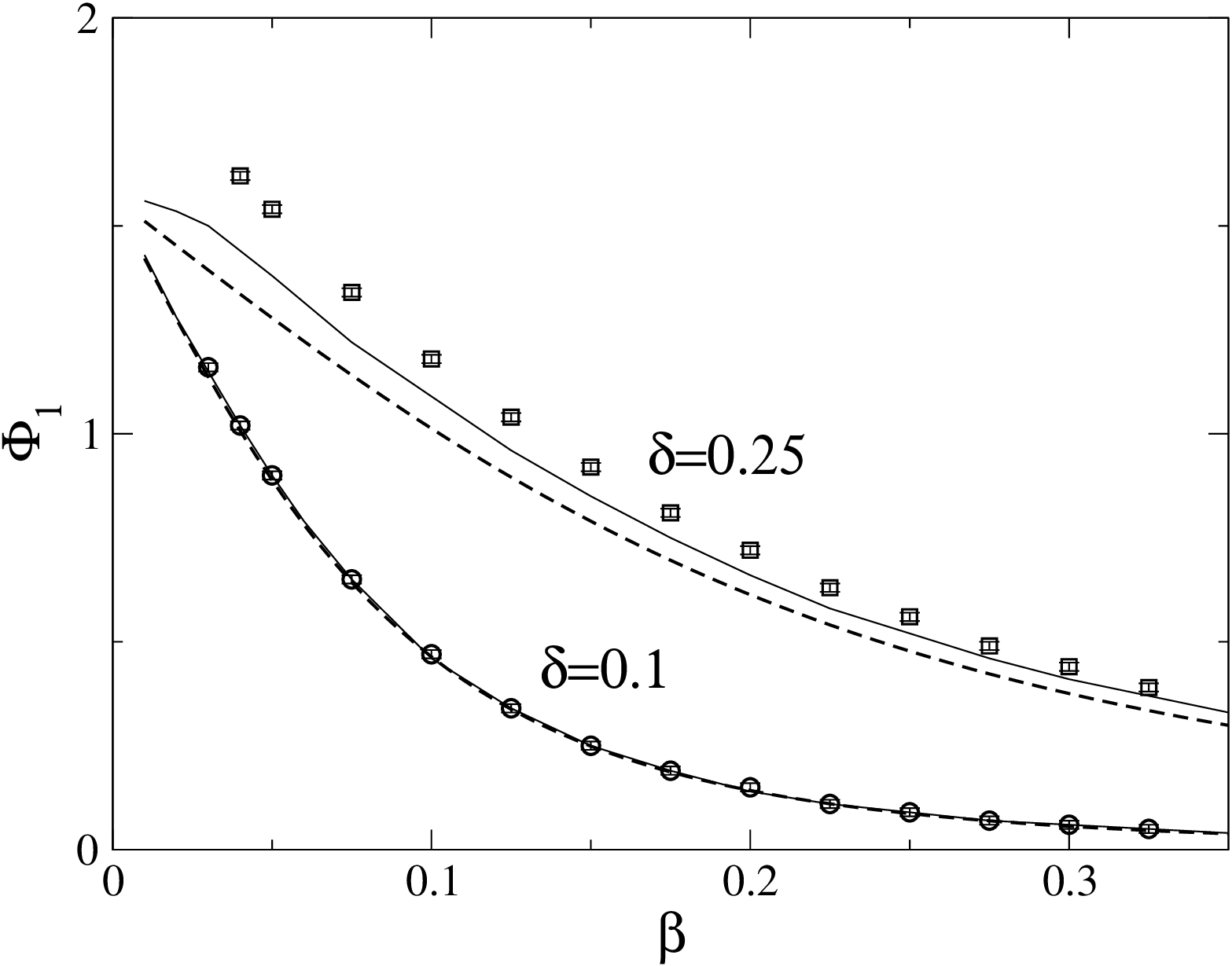}
 \end{tabular}
\end{center}
\caption{Dependence of the function $\Phi_{1}=-\theta_{2}^{c}$ on
the dissipation coefficient for $m=2$ and $\beta \ge 0.01$.
In the simulations as well
as in the CC, we find $\theta_{2}^{c}$ as the value of the phase
between $-\pi$ and 0 for which the motion is suppressed. Solid
line: CC result; dashed line: Eq.\ (\ref{fase-damp}); symbols:
numerical simulations of the full sine-Gordon equation. Parameters
are $\epsilon_1=\epsilon_2=0.2$ and $\theta_{1}=0$.}
\label{damping}
\end{figure}

For the case $m=3$, the shift-symmetry is not broken and therefore
ratchet motion can not take place. Indeed, the calculation of the
average velocity based on the CC approach gives zero for all
orders of the expansion. For this case the frequencies of the ac
force (or the momentum) are odd harmonics of $\delta$ ($\delta$
and $3\delta$), whereas only even harmonics of $\delta$ are found
in the kink width oscillations  ($2n\delta$, $n \in N$). The
complete selection rules for $m=2,3,4$ appear in Table \ref{tab1},
where for a given $m$ one identifies the harmonics corresponding
to the oscillations of the width of the kink. Resonances take then
place when at least one of these harmonics coincides with one of
the two harmonics of the ac force. We can verify that if this
condition is fullfiled, the shift symmetry is broken and
viceversa. In the simplest case, $m=1$, for which  the driving
consists only of a single harmonic, the ratchet effect is also
absent. Looking at Eq.\ (\ref{2.6}) it is immediately realized
that only terms with frequency $2\delta$ appear in the equation
(\ref{cc-l}), so that the resonance condition is not achieved in
agreement with previous results \cite{niurka2}. This discussion in
terms of resonant frequencies can be extended, in an analogous but
cumbersome calculation, to any positive integer number of the
frequency of the second harmonic, i.e., for higher integer values
of $m$. Finally, an interesting remark is that the resonance
condition can also be interpreted as a synchronization between the
ac force and the oscillations of the width of the kink, i.e.,
directional transport occurs whenever the ac force is locked to
the oscillations of the width \cite{salerno2}.

\subsection{Mechanism for directional motion}

In the previous subsection we have established the conditions for
net motion of the kink to take place. We argued that the fact that
the width of the kink  is a dynamical variable is not a sufficient
condition for the existence of directed motion and, on the
contrary, net motion can arise only when, first, at least one of
the two harmonics of the ac driven force  appears in the
oscillations of the width of the kink (which is equivalent to
breaking the shift symmetry) and, second, when the time-reversal
symmetry is broken. However, these two conditions, albeit simply
stated, do not allow for a physical or intuitive understanding of
the mechanisms at work in the system. In order to shed light on
those mechanisms, it is convenient to write Eq.\ (\ref{cc-p}) as
\begin{equation}
\ddot{X}+\beta \dot{X}-
\frac{\dot{X}\dot{l}}{l}=-\frac{ql(t)f(t)}{
M_{0}l_{0}}.\label{cc-X}
\end{equation}
Cast in this form, it is easy to
realize the existence of an  effective driving force, given by
\begin{equation}
\label{nueva}
f_{ef}=-\frac{\textstyle ql(t)f(t)}{\textstyle M_{0}l_{0}}
\end{equation}
whose amplitude is modulated by the oscillations of the width. The
consequence is that, while the average of the external ac force is
always zero, the average of this effective force term vanishes
when the harmonics of the external driving force $f(t)$ do not
match those in the oscillations of the width or when they
oscillate with at least one common frequency but with appropriate
critical phases. Such a nonzero effective driving force is in fact
the same as having a dc force acting on the kink. This means that
the energy is {\em inhomogeneously} pumped into the system
\cite{flach2}, through a process that is modulated by the kink
width oscillations.

On the other hand, Eq.\ (\ref{cc-X}) has a term $\dot{X}\dot{l}/l$
that describes the energy transfer from the traslational mode to
the internal mode. Instead of analyzing which portion of energy
delivered by the ac force is used for the dynamics of the internal
mode, we find it more illuminating to study how different this
transfer is in the two halves of one period. To this end, Fig.\
\ref{effec-force}(a) presents the effective force for the case
$m=2$ for certain parameters. The difference between the brown and
grey areas divided by the period represents the average of the
effective force in one period. The computation shows that $\langle
f_{ef}\rangle =0.0055$. Consequently, the center of the kink
should move in the positive direction and, indeed, as is depicted
in Fig.\ \ref{effec-force}(b), such motion takes place. In
addition, the mean velocity is $\langle v\rangle=0.108$, the kink
moves with constant velocity on average, and hence the average
acceleration is zero. One can check that $\beta \, \langle
v\rangle=0.0054$ in close agreement with the result found above
for the effective force. We also computed the average of the
coupling term which gives $\langle \dot{X}\dot{l}/l
\rangle=-0.0001$. This tells us that, on average, the coupling
term is equivalent to a friction force but also that its value is
much smaller compared to the average of the driving force. Hence,
most of the net force is used in the traslational motion.

\section{Numerical simulation results}

The analytical results of the previous section were derived within
the CC approach, which amounts to consider that perturbations
affect only the $X$ and $l$ variables of the kink but not its
shape, i.e., phonons have been neglected. Therefore, we expect
that the results will be valid for small amplitudes and
frequencies of the ac force, and also not too small damping
coefficient. In any event, they must be compared to numerical
simulations of the full partial differential equations (\ref{sG})
and (\ref{phi4}). To this end, we have integrated numerically both
equations, using the Strauss V\'azquez scheme \cite{vaz} as well
as a fourth-order Runge-Kutta method, choosing a total length of
$L=100, 300$, with steps $\Delta t=0.01$,  $\Delta x=0.1$.  We
have used aperiodic boundary conditions with a kink at rest as initial
condition. In the following we will monitor the dynamics induced
by the ratchet effect through the dependence of the average
velocity on the phases of the bi-harmonic force and the
dissipation coefficient, which in turn requires the computation of
the center and the width of the kink from the simulations of the
full equations.

For the numerical calculation of the center and width of the kink
we have improved the procedure suggested in \cite{buscar}, taking
into account the oscillations of the ground state due to the
action of the ac driving. In particular, for the sG kink (and for
the $\phi^4$ kink, changing parameters as indicated), this method
reduces to searching, for a given and fixed time, in the discrete
lattice, those points $x_{n}$ and $x_{n+1}$ such that $\phi_{n}
\le \pi +\phi_{\mbox{vac}}(t)$ and $\phi_{n+1}\ge
\pi+\phi_{\mbox{vac}}(t)$, where $\phi_{n}=\phi(x_{n},t)$,
$\phi_{\mbox{vac}}(t)=\phi(x\to\pm\infty,t)$ represents the
oscillations of the background field. In this case this function
can be computed as $\phi_{N}-2\pi$ ($\phi_N-1$ for $\phi^4$) where
$N=L/\Delta x$, $L$ being the total length of the numerically
simulated system. Subsequently we estimate, by using a linear
interpolation method with two points $(x_{n},\phi_{n})$ and
$(x_{n+1},\phi_{n+1})$, the corresponding point $\tilde x_{n}$ (it
will be our computed center of the kink $X(t)$) where
$\phi=\pi+\phi_{\mbox{vac}}$ ($\phi=\phi_{\mbox{vac}}$ for
$\phi^4$). Second, in order to compute the width of the kink, we
look for the value of $l(t)$ that minimizes the expression
\begin{equation}
\label{nueva2} \sum_{k=1}^N
\left|\phi_{k}-\left(\phi_{0}\left[\frac{\textstyle k \,\Delta
x-\tilde x_{n}}{\textstyle l(t)}\right]
+\phi_{\mbox{vac}}\right)\right|^2,
\end{equation}
where $\phi_{0}$ is defined by using (\ref{RicesG}) (or
(\ref{Ricephi4}) in the case of the $\phi^4$ system) and $k=1,2...n...N$.

In the following subsections, we will present the numerical
analysis in two parts according to the two models we deal with,
namely sG and $\phi^4$. In each system we will analyze the ratchet
dynamics of the kink for $m=2, 4$ and the zero average velocity
for $m=3$.
\begin{figure}
\begin{center}
\begin{tabular}{cc}
\hspace{-1.5cm}
\includegraphics[width=7cm,height=7cm]
{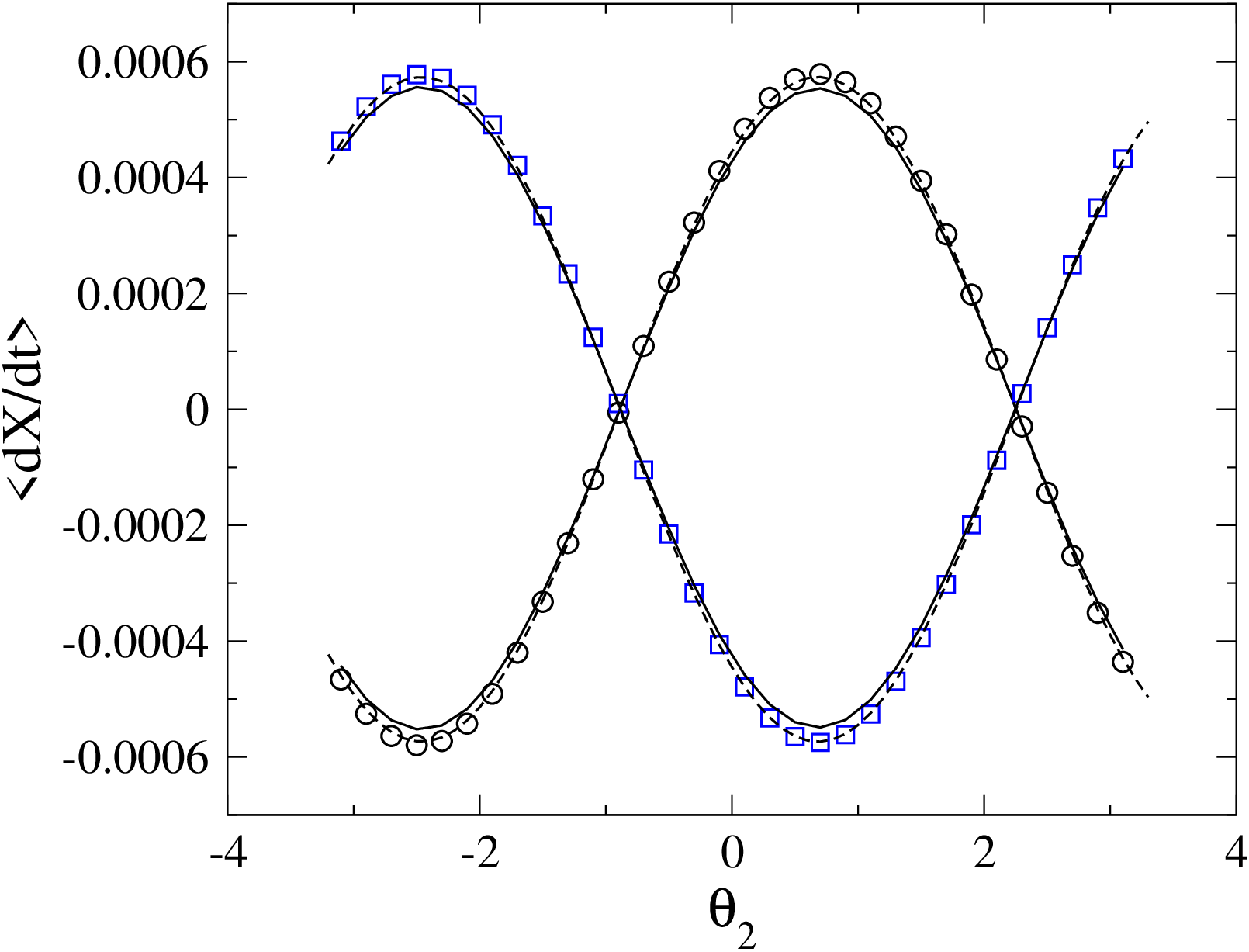}
\includegraphics[width=6.5cm,height=7cm]
{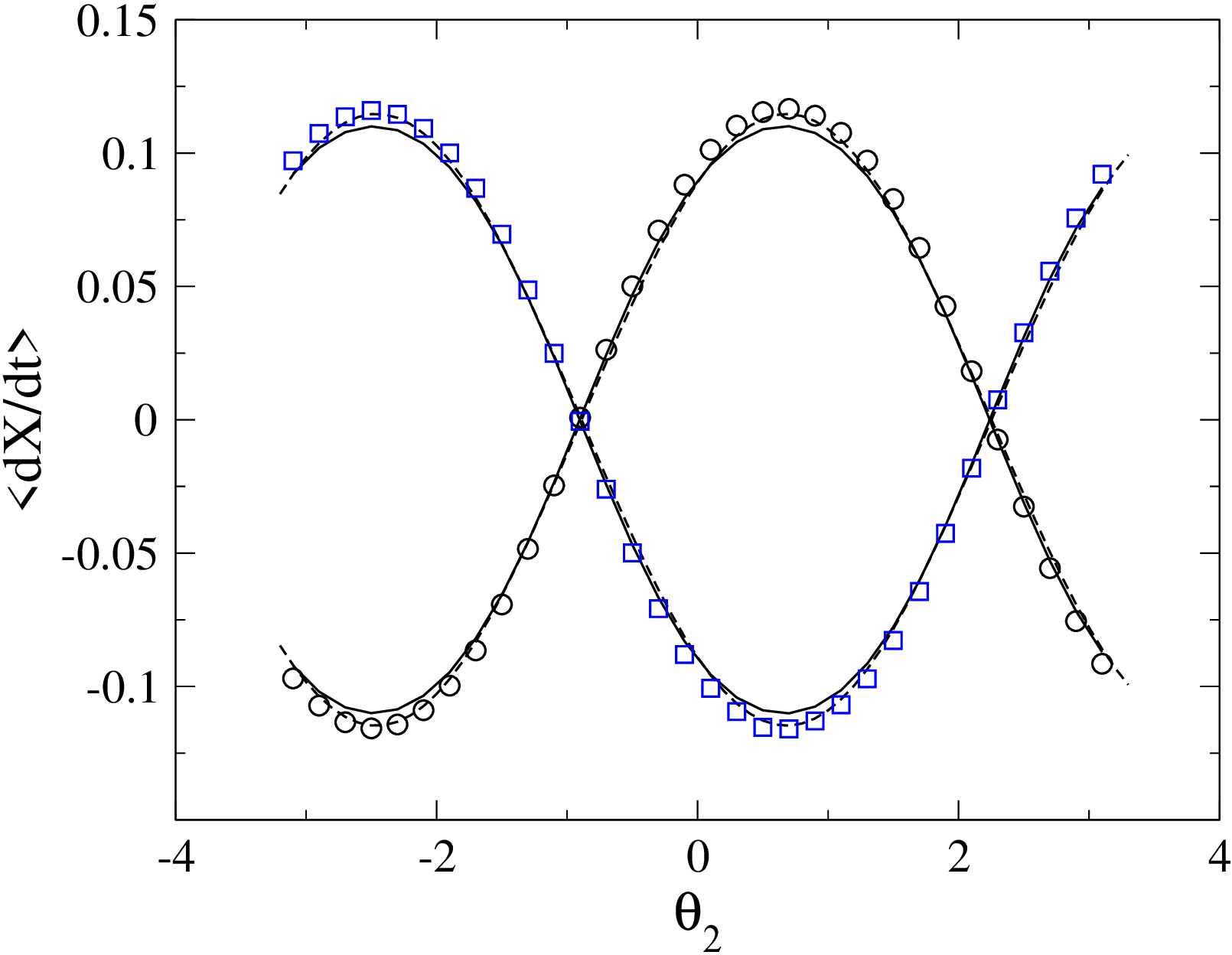}
\end{tabular}
\end{center}
\caption{Dependence of the average velocity on $\theta_{2}$ for
$m=2$. Left panel: Parameters are $\epsilon_1=\epsilon_2=0.02$.
Right panel: $\epsilon_1=\epsilon_2=0.2$. Other parameters are
$\beta=0.05$, $\delta=0.1$. In both panels two values for
$\theta_{1}$ are considered: blue squares are the results of the
numerical simulations for $\theta_{1}=0$, and black circles
correspond to $\theta_{1}=\pi/2$. Solid lines correspond to the
numerical solution of the CC Eqs.\ (\ref{cc-p})-(\ref{cc-l}). In
the left (right) panel dashed lines correspond to
$\epsilon\langle\dot{X}_{1}\rangle$
($\epsilon\langle\dot{X}_{1}\rangle/5$), obtained from Eq.\
(\ref{2.16}).} \label{Fig1-m2}
\end{figure}
\begin{figure}
\begin{center}
\includegraphics[width=6.5cm,height=7cm]{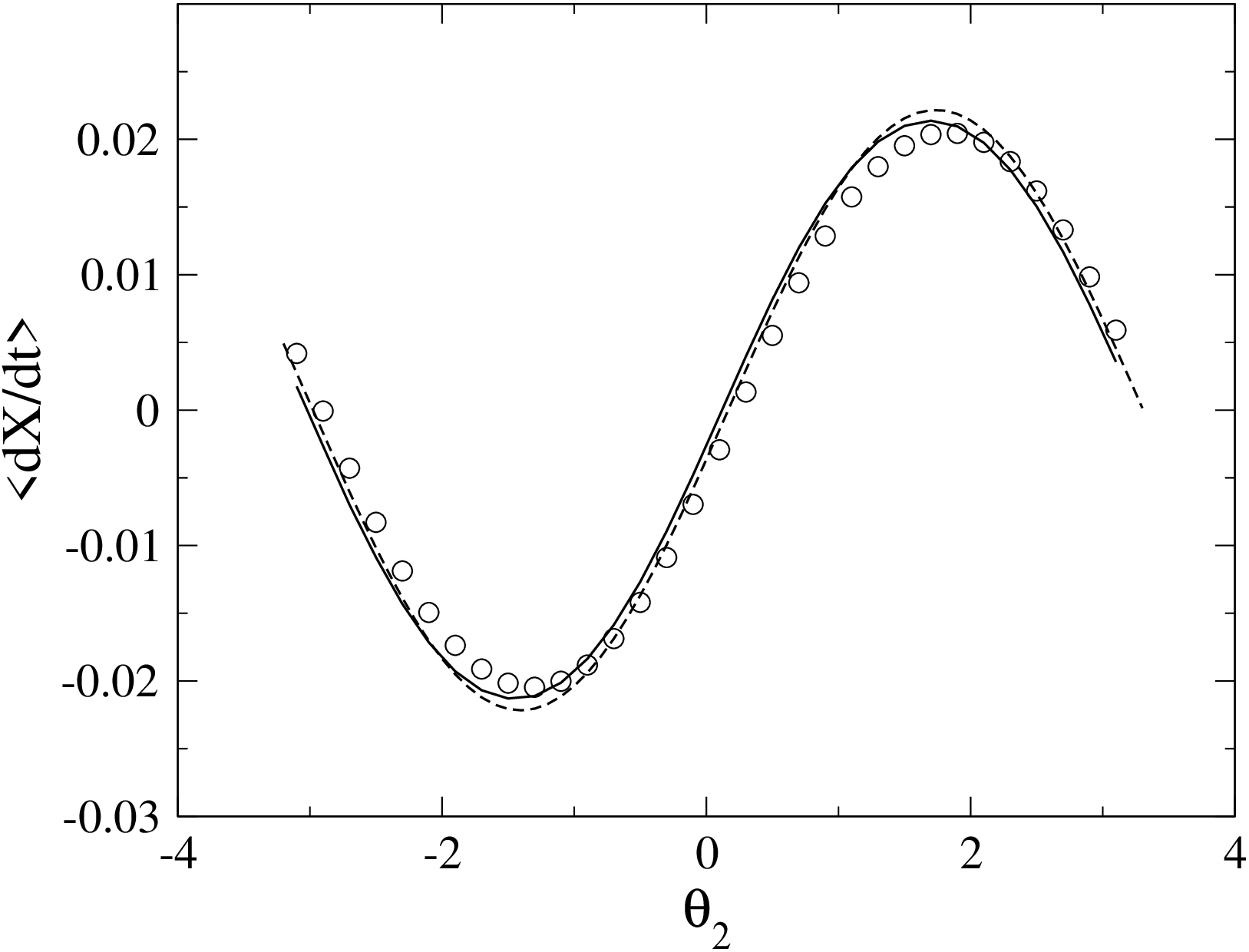}
\end{center}
\caption{Dependence of the average velocity on $\theta_{2}$ for
$m=4$. Circles correspond to numerical simulations; solid line to
the numerical solution of CC Eqs.\ (\ref{cc-p})-(\ref{cc-l}), and
dashed line stands for $\epsilon^2\langle\dot{X}_{2}\rangle/9$.
Parameters are $\epsilon_1=\epsilon_2=0.2$, $\theta_{1}=0$,
$\delta=0.1$ and $\beta=0.05$.} \label{sG-cc-pde-m4}
\end{figure}
\begin{figure}
\begin{center}
\begin{tabular}{lc}
\includegraphics[width=7.5cm,height=7.cm]{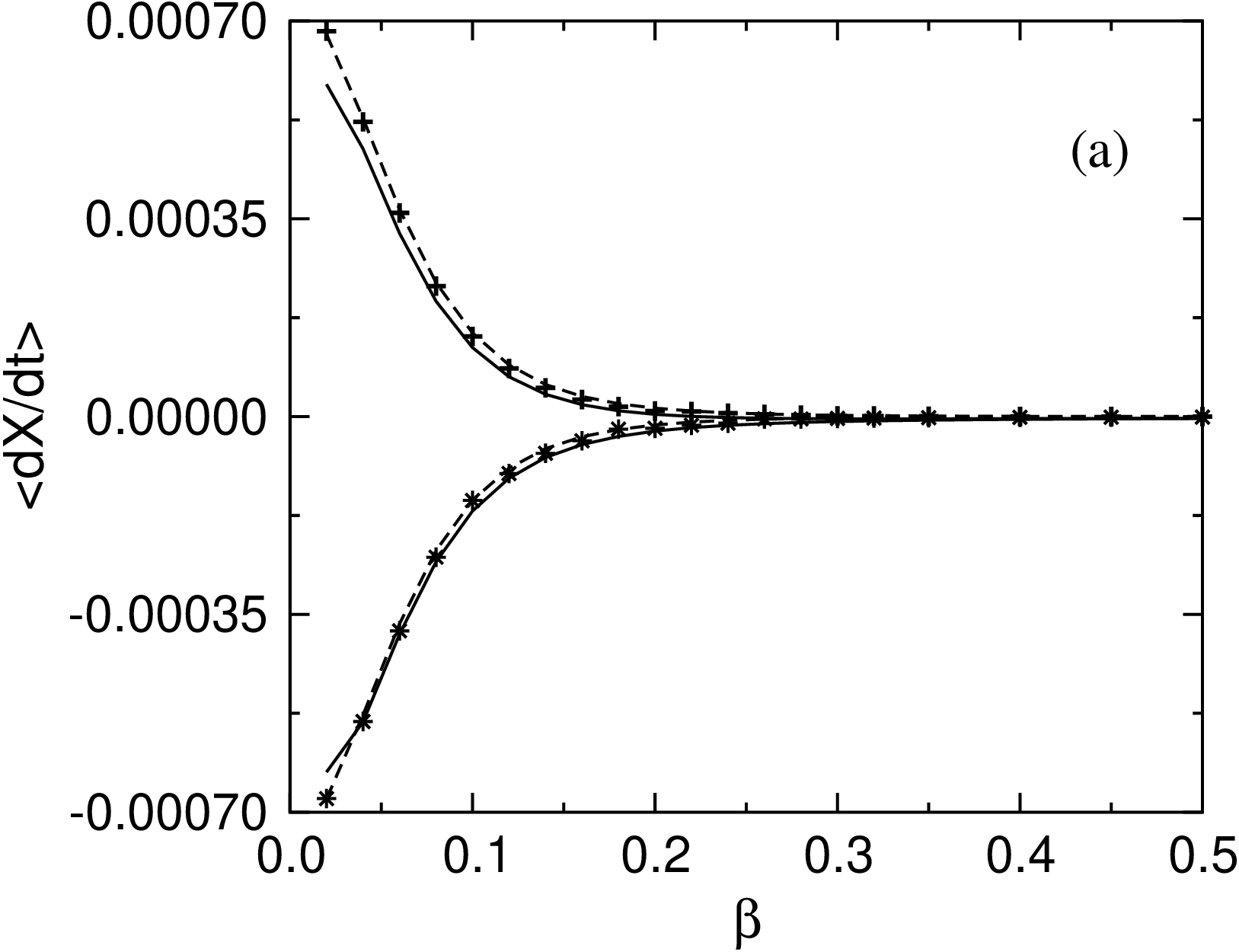}\hspace{3mm}
\includegraphics[width=7.5cm,height=7.cm]{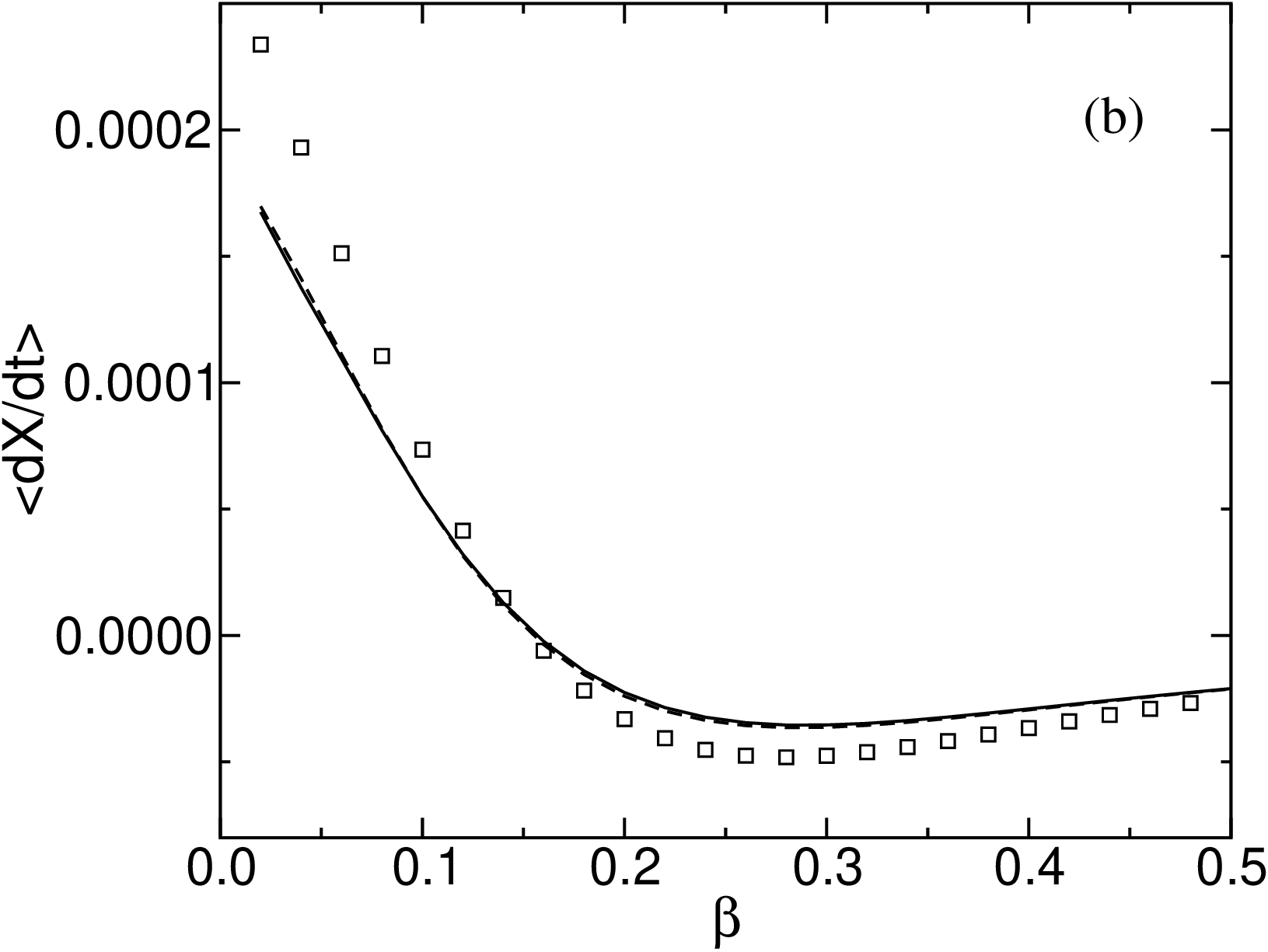}
\vspace{0.4cm}
\end{tabular}
\\
\includegraphics[width=8.cm,height=7.cm]{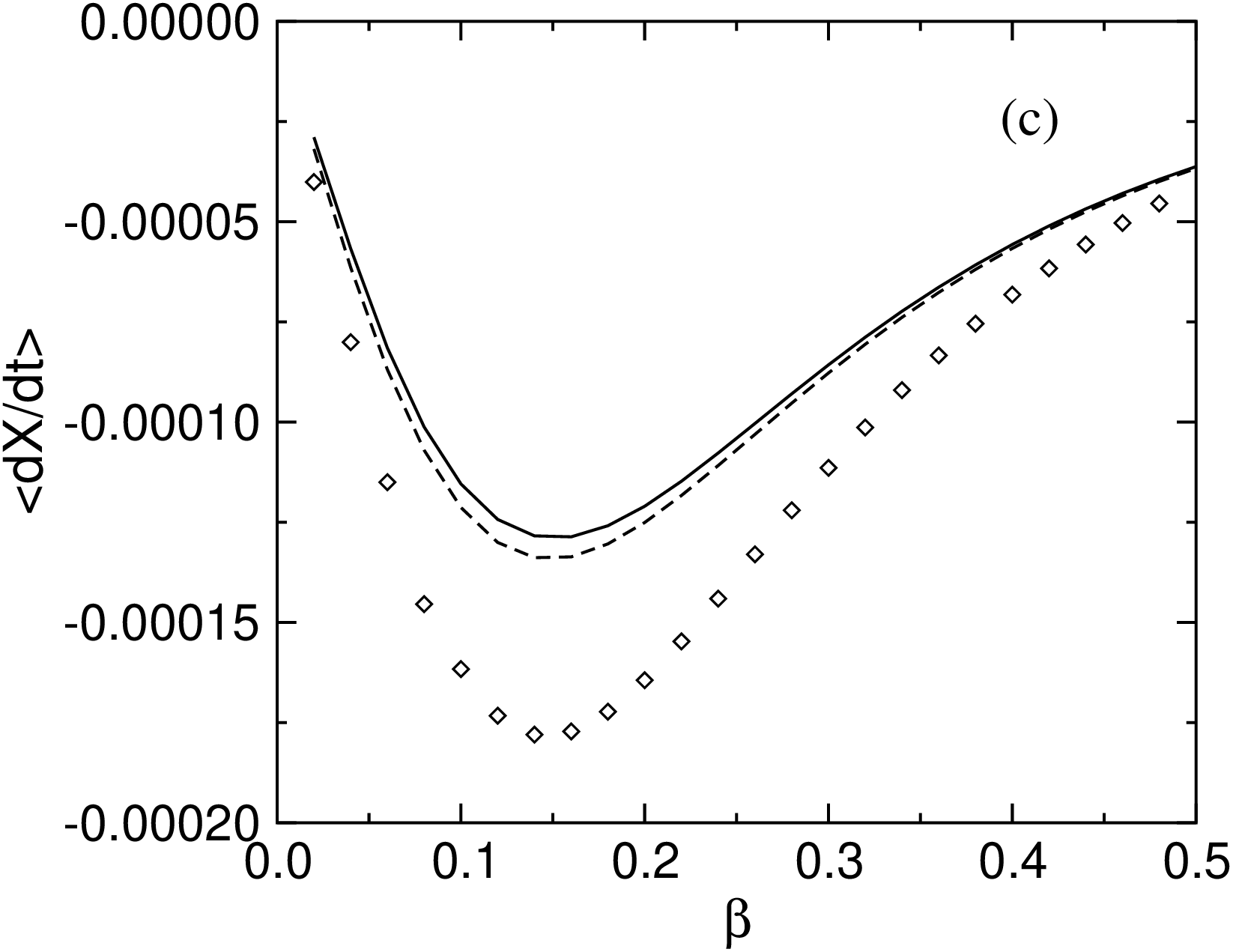}
\end{center}
\caption{Average velocity of a sG kink as a function of the
dissipation for $\beta \ge 0.02$. Panel (a): $\epsilon_{1}=\epsilon_{2}=0.02$,
$\delta=0.1$, $\theta_{1}=\pi/2$; Numerical simulations:
$\theta_{2}=\pi$ ($\ast$); $\theta_{2}=0$ (+). Panel (b):
$\epsilon_{1}=0.04$, $\epsilon_{2}=0.026$, $\delta=0.25$,
$\theta_{1}=-\pi/2$, $\theta_{2}=-\pi/2+0.8$; Numerical
simulations: ($\square$). Panel (c): $\epsilon_{1}=0.04$,
$\epsilon_{2}=0.026$, $\delta=0.25$, $\theta_{1}=-\pi/2$,
$\theta_{2}=-\pi/2$; Numerical simulations: ($\lozenge$). In each
figure the solid line shows $\langle V \rangle$ obtained from the
numerical solution of  Eqs.\ (\ref{cc-p})-(\ref{cc-l}), and the
dashed line is $\langle V \rangle$ calculated from Eq.\
(\ref{2.16}).} \label{sG-beta}
\end{figure}
\begin{figure}
\begin{center}
\includegraphics[width=8cm,height=7.cm]{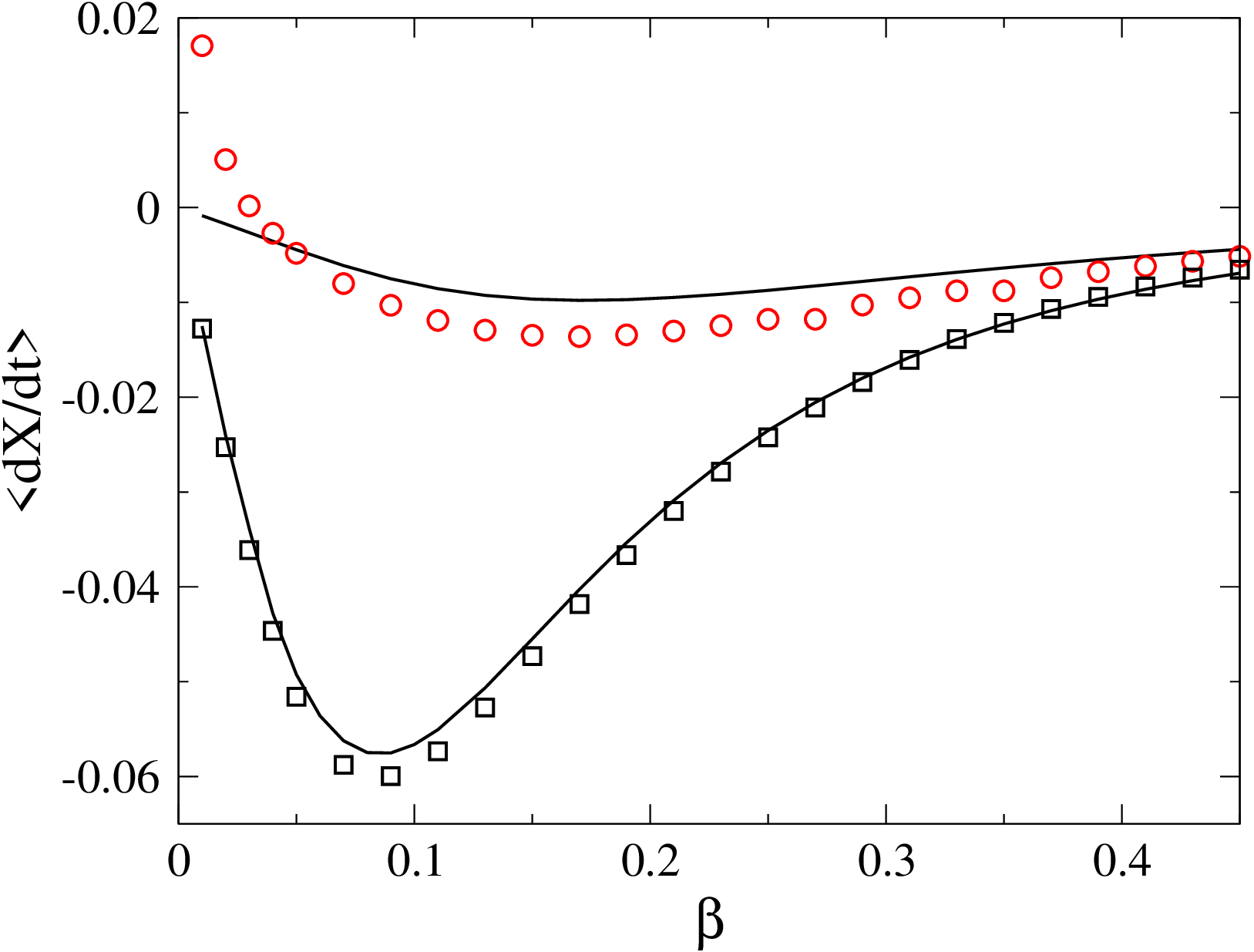}
\end{center}
\caption{Average velocity of a sG kink as a function of the
dissipation for  $0.01 \le \beta \le 0.45$
and for two different frequencies: $\delta=0.1$ (squares);
$\delta=0.25$ (circles). The other parameters are
$\epsilon_{1}=0.2$, $\epsilon_{2}=0.12$, $\theta_{1}=-\pi/2$,
$\theta_{2}=-\pi/2$. In both cases the solid lines show the
results obtained from the numerical solution of Eqs.\
(\ref{cc-p})-(\ref{cc-l}). } \label{sG-beta2}
\end{figure}

\subsection{Sine-Gordon model}

Our first comparison between the CC analyisis and the numerical
simulations is presented in Fig.\ \ref{effec-force}(b), where the
time evolution of the center of the kink is presented. As can be
seen from the plot, the agreement between our analytical
approximation and the numerical results is excellent, even for
long times, suggesting that we can be confident that the CC
approach captures the ratchet dynamics of the system.
Nevertheless, to complete the analysis and fully confirm our
analytical predictions, it is crucial to understand how the width
of the kink evolves when it is driven by the bi-harmonic external
force. The evolution in time of the width of the kink as obtained
from the numerical simulations of the full sG equation and from
the CC equations are compared in the left panel of Fig.
\ref{width-mixing}. Once again, we observe an excellent agreement
between simulations and the CC framework. In order to validate our
predictions about the resonance criterion, we proceed to the
determination of the Fourier components for the oscillations of
the width of the kink (see right panel of Fig.
\ref{width-mixing}). The Discrete Fourier Transform (DFT) shows an
excelent agreement between the numerical simulations of Eq.\
(\ref{sG}) and the numerical solutions of Eqs.\
(\ref{cc-p}-\ref{cc-l}), confirming the resonance criterion. For
instance, we can observe that, for $m=2$, the spectrum of $l(t)$
contains the frequencies $\delta$ and $2\delta$, the same
frequencies of the bi-harmonic external force. Consequently, one
could expect a net motion of the soliton, as was pointed out in
Table \ref{tab1} and as the numerical simulations show. To further
confirm the resonance criterion, we analyze the spectrum of $l(t)$
for $m=3, 4$, plotted in Fig.\ \ref{fig:dftsG}. For $m=3$ neither
$\delta$ nor $3\delta$ appear in the spectrum, thus confirming our
predictions (see Table \ref{tab1}): We observe only the presence
of even harmonics, and hence we expect an oscillatory motion of
the center of the kink similar to that obtained for a force with
only one harmonic component \cite{niurka5}. Conversely, for $m=4$
we observe the appearance of frequencies $\delta$ and $4\delta$.
Therefore, the occurrence of a net motion is also expected as in
case of $m=2$ and is indeed confirmed by the simulations (not
shown).

Having verified our resonance criterion for the ocurrence of
ratchet phenomena, we now turn to discussing how the average
velocity depends on the parameters of the ac force and damping. In
Fig.\ \ref{sGm2m3m4} we plot the average velocity of the kink,
computed from the numerical simulations of the full PDE, for $m=2,
3$, and $4$ versus the phase of the second harmonic $\theta_2$
[see Eq.\ (\ref{ac})]. In this plot, the predictions for the
existence or not of motion for different values of $m$ are
confirmed. Notice the sinusoidal dependence of the mean velocity
on the phase, as predicted by the expressions (\ref{2.16}) and
(\ref{2.17}). Another relevant feature of Fig.\ \ref{sGm2m3m4} is
the difference of the mean velocity between the cases $m=2$ and
$m=4$. In principle, net motion can occur for any even value of
$m$; however, we observe that the maximum of the velocity
decreases when $m$ is increased. This can be understood from Eqs.\
(\ref{2.16a}) and (\ref{2.17a}), which predict that $\langle V
\rangle \sim \epsilon_{1}^2 \epsilon_{2}$ for $m=2$ and $\langle V
\rangle \sim \epsilon_{1}^4 \epsilon_{2}$ for $m=4$. For greater
values of $m$, $\langle V \rangle \sim \epsilon_{1}^m
\epsilon_{2}$ \cite{ricardo}, so if $\epsilon_{1}$ and
$\epsilon_{2}$ are small enough, the velocity goes to zero as $m$
is increased. Furthermore, the amplitudes of the peaks of the DFTs
corresponding to the frequencies $\delta$ and $2\delta$ for $m=2$
are greater than their counterparts for $m=4$, $\delta$ and $4
\delta$ (see right panel of Figs. \ref{width-mixing} and
\ref{fig:dftsG}). In this figure we also notice that there are
some values of the phase $\theta_{2}$ for which there is no net
motion: For example, for $m=2$, these values are approximately
given by $\theta_{2}^{c}=-0.9+n \pi$ ($n=0, 1, ...$). We will
refer to those values of the phase for which there is no net
motion as critical values. For $m=2$, the whole set of critical
phase values as a function of dissipation and frequencies is
provided by the expression
$\theta_{2}^{c}=-\Phi_{1}(\beta,\delta)+n\pi$. These correspond to
those values of the phase for which the condition (\ref{src}) is
not fulfilled.

The dependence of $\Phi_{1}(\beta,\delta)$ on the dissipation
coefficient and the frequency of the ac force is shown in Fig.\
\ref{damping}.
The figure shows that for the largest value of the frequency, the
agreeement between the CC theory and the simulations of the full
partial differential equation is only qualitative. The reason for
this lies in the fact that, when the harmonics of the ac force
contain the frequencies $\delta$ and $2\delta$, the width of the
kink is excited with frequencies $2\delta$ and $4\delta$,
respectively (because of the term $P^2$ in (\ref{cc-l}), see also
previous studies \cite{niurka2,niurka3}). If we chose
$\delta=0.25$, $l(t)$ will oscillate with $2\delta=0.5$,
$4\delta=1$, or even higher frequencies, which are inside the
phonon band, which lies above unit frequency. Therefore, for large
enough amplitude of the force and small dissipation, strong
excitation of phonons is expected (even kink-antinkinks can
appear) and correspondingly the failure of our CC approach, since
it does not take into account the phonon contribution to the
motion. This is yet another reason why simulations for very low
values of the dissipation coefficient (and hence with larger
phonon production) are not considered in the plot.

Let us now concentrate on the $m=2$ case and compare the average
velocities as a function of $\theta_{2}$, computed from the direct
numerical simulations of Eq.\ (\ref{sG}) and from the numerical
solution of the CC equations (\ref{cc-p})-(\ref{cc-l}). In Fig.\
\ref{Fig1-m2} we observe an excellent agreement between these two
average velocities. In the left panel, overimposed to these
values, we show also the approximate values of $\langle V \rangle$
obtained from Eq.\ (\ref{2.16}). In the right  panel, the
amplitudes of the two harmonics have been increased by one order
of magnitude; this leads to noticeable phonon production in the
simulations and consequently to a quantitative disagreement
between the simulations and the CC approach
, and therefore a factor $1/5$ has been introduced in order to
adjust the values of $\langle V \rangle$ obtained from Eq.\
(\ref{2.16}) to the results of the numerical simulations and the
numerical solution of CC equations (as is obvious, the qualitative
agreement is perfect and in particular for the critical phase
values the agreement is quantitative). For the case $m=4$ the
situation is the same: There is good agreement between the results
of the CC equations and the simulations, although once again the
analytical expression $\langle V \rangle$ fits only qualitatively
to the numerical results (see Fig.\ \ref{sG-cc-pde-m4}).

One important point, relevant in experimental contexts, is the
prediction of the CC theory about the dependence of the mean
velocity on the dissipation coefficient. According to the standard
behavior of point particles under friction, one would expect a
monotonic decreasing of the average velocity to zero as a function
of the damping. However, the existence of an optimal damping for
the occurrence of net motion, and the appeareance of current
reversal upon varying the damping (see \cite{Salerno-mixing},
\cite{salerno2}), show a richer and more complex behaviour in the
case of the soliton ratchets we are discussing, depending on the
parameters of the ac force. These phenomena, expected also from
Eqs.\ (\ref{2.16}) and (\ref{2.17}), are shown in Fig.\
\ref{sG-beta}, where for the first sets of parameters (Fig.\
\ref{sG-beta}a) we observe that the velocity drastically decreases
to zero as $\beta$ is increased. Notice that in this case the
time-reversal symmetry of the ac force is broken in the optimal
way \cite{ricardo} by the given phases. In Fig.\ \ref{sG-beta}c,
when the phases are chosen in a manner such that the time reversal
symmetry $f(t)=f(-t)$ is fulfilled in the absence of dissipation,
an optimal value of $\beta$ for the transport is observed. In this
case, when the dissipation coefficient increases starting from
$\beta=0$, the time-reversal symmetry begins to break, and
therefore an increase of the average velocity is expected. On the
other hand, increasing the damping tends to suppress any motion.
Hence, in between these two opposite mechanisms an optimal value
of the damping must appear \cite{ricardo}. Between these two limit
cases, we can see the appeareance of current reversal (see Fig.\
\ref{sG-beta}b) as was pointed out in \cite{Salerno-mixing}.

From Fig. \ref{sG-beta} we can see that the best agreement between
CC theory and numerical simulations is obtained for small values
of the frequencies of the ac force. Indeed, in Fig.\
\ref{sG-beta}c, for $\delta=0.25$, due to the phonon contribution
(see the discussion above), we observe only a qualitative
agreement between the numerical simulations and the CC theory.
This disagremeent is even clearer in Fig.\ \ref{sG-beta2}, where
we take large values of the amplitudes of the ac force and compare
the mobility of the kink for  $\delta=0.1$ and $\delta=0.25$. We
can observe that for the largest value of frequency together with
the smallest values of $\beta$ the results from the theory do not
fit quantitatively the values obtained from the numerical
simulations. Furthermore, we even observe current reversal in the
results of the numerical simulations, contrary to the predictions
of the CC theory. As we explained above, this disagreement arises
from the large production of phonons for this value of the
frequency, unaccounted for in the CC approach.
\begin{figure}
\begin{center}
\begin{tabular}{lc}
\hspace{-1cm}
\includegraphics[width=7.5cm,height=7.cm]{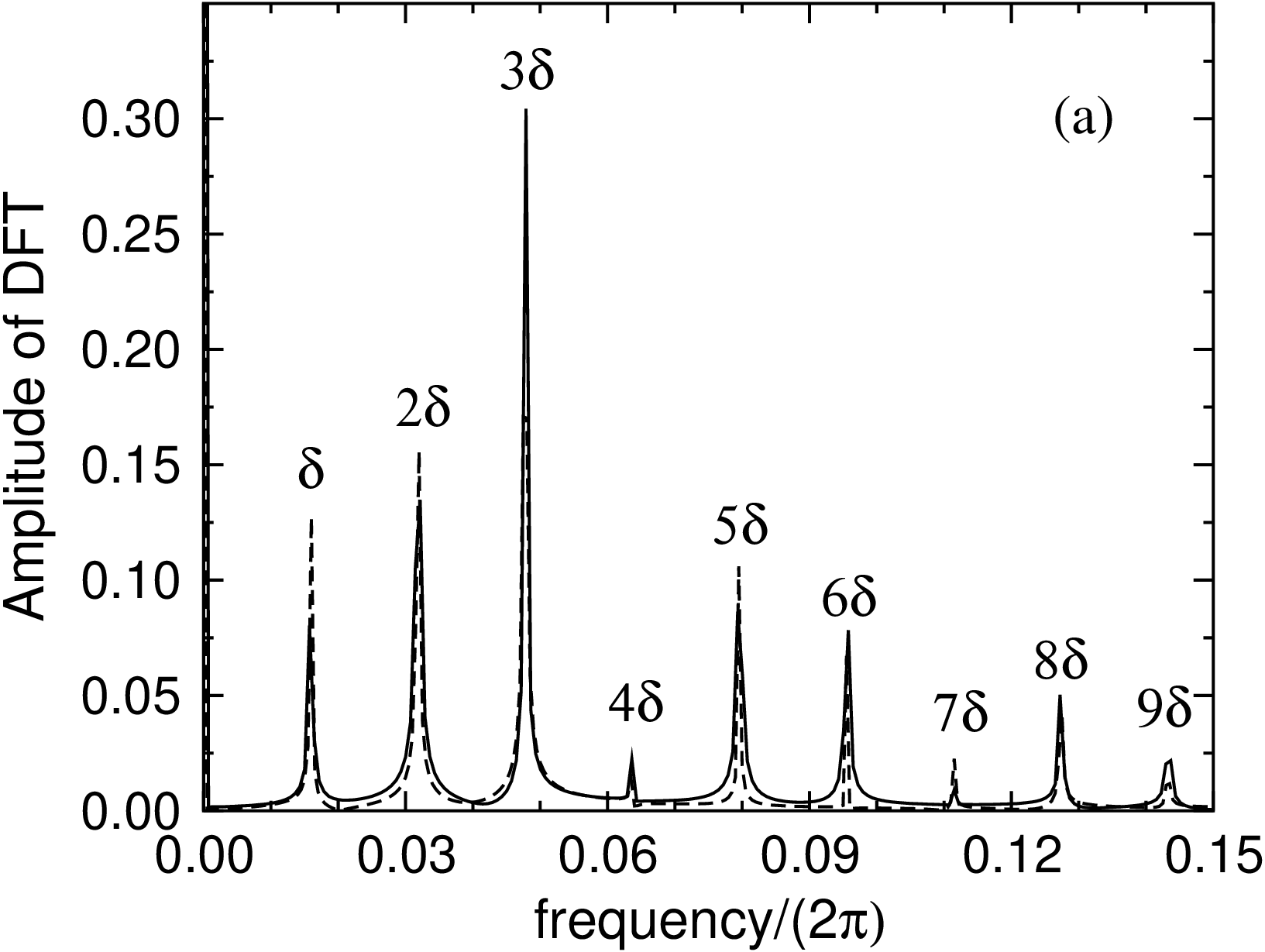}
\includegraphics[width=7.5cm,height=7.cm]{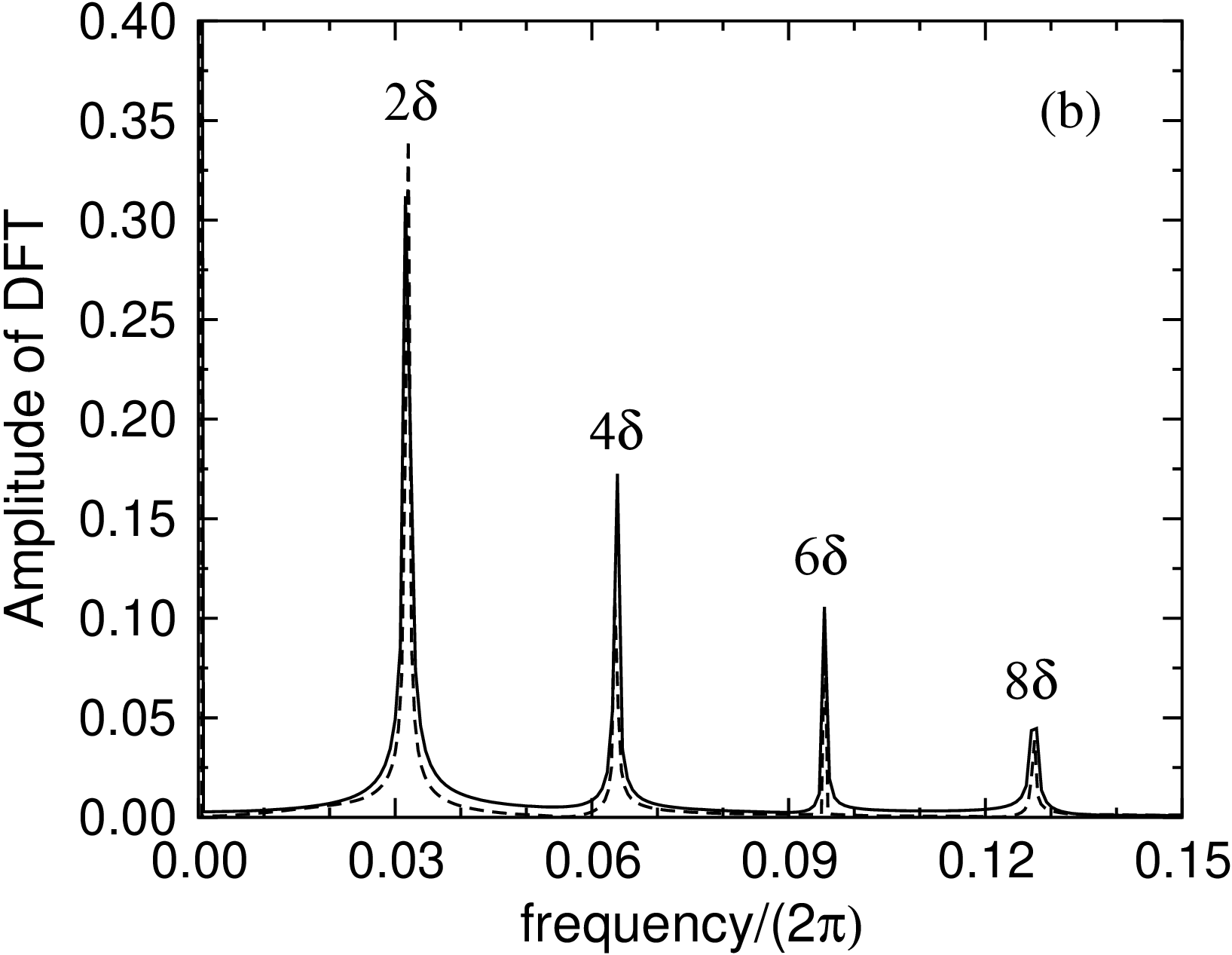}
\vspace{0.8cm}
\end{tabular}
\includegraphics[width=7.5cm,height=7.cm]{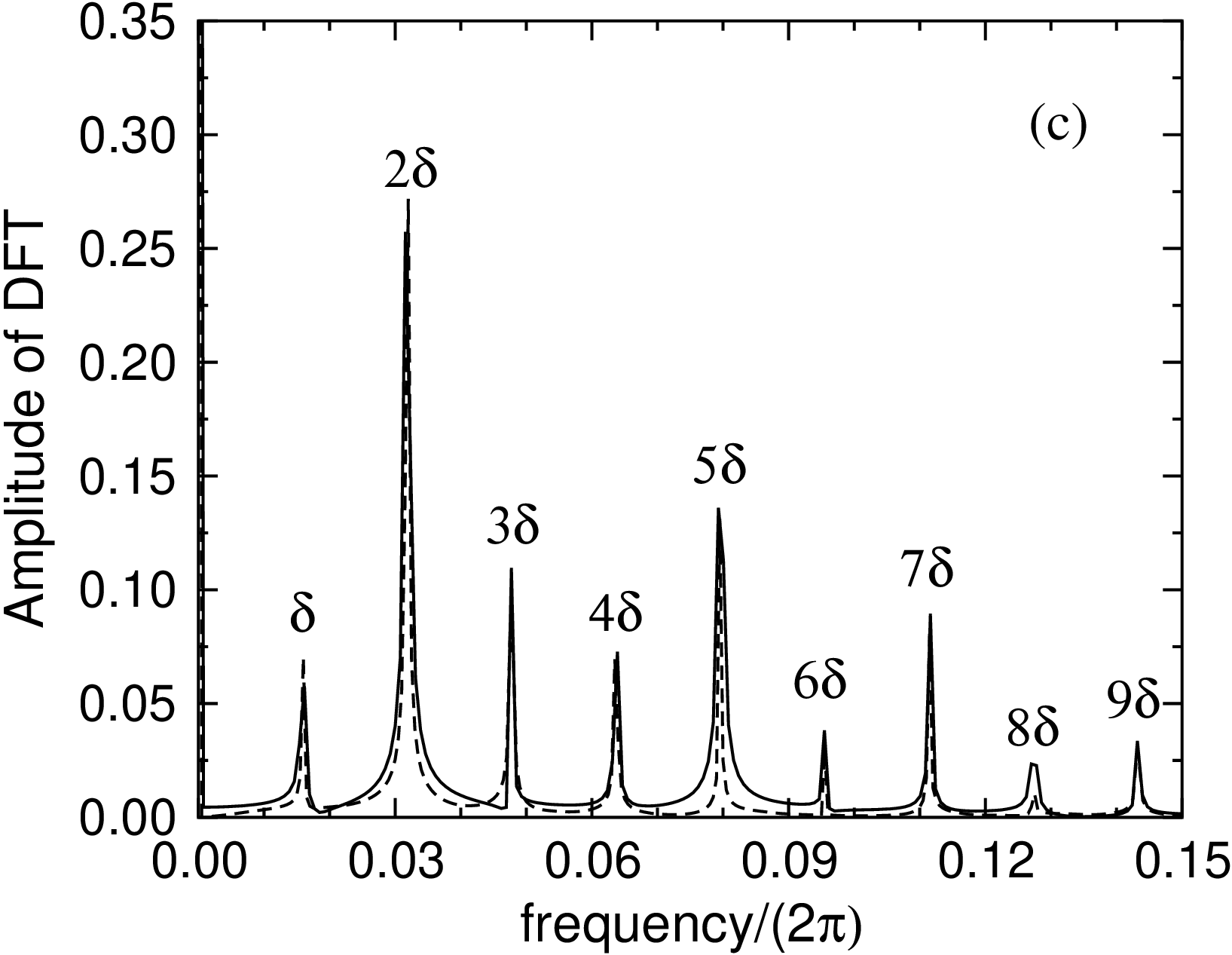}
\end{center}
\caption{Discrete Fourier Transform of the width of a $\phi^4$
kink. (a): $m=2$; (b): $m=3$; (c): $m=4$. Solid line: Simulations
of Eq.\ (\ref{phi4}). Dashed line: numerical integration of the CC
equations.} \label{dftmphi4}
\end{figure}
\begin{figure}
\begin{center}
\begin{tabular}{lc}
\includegraphics[width=7.5cm,height=7cm]{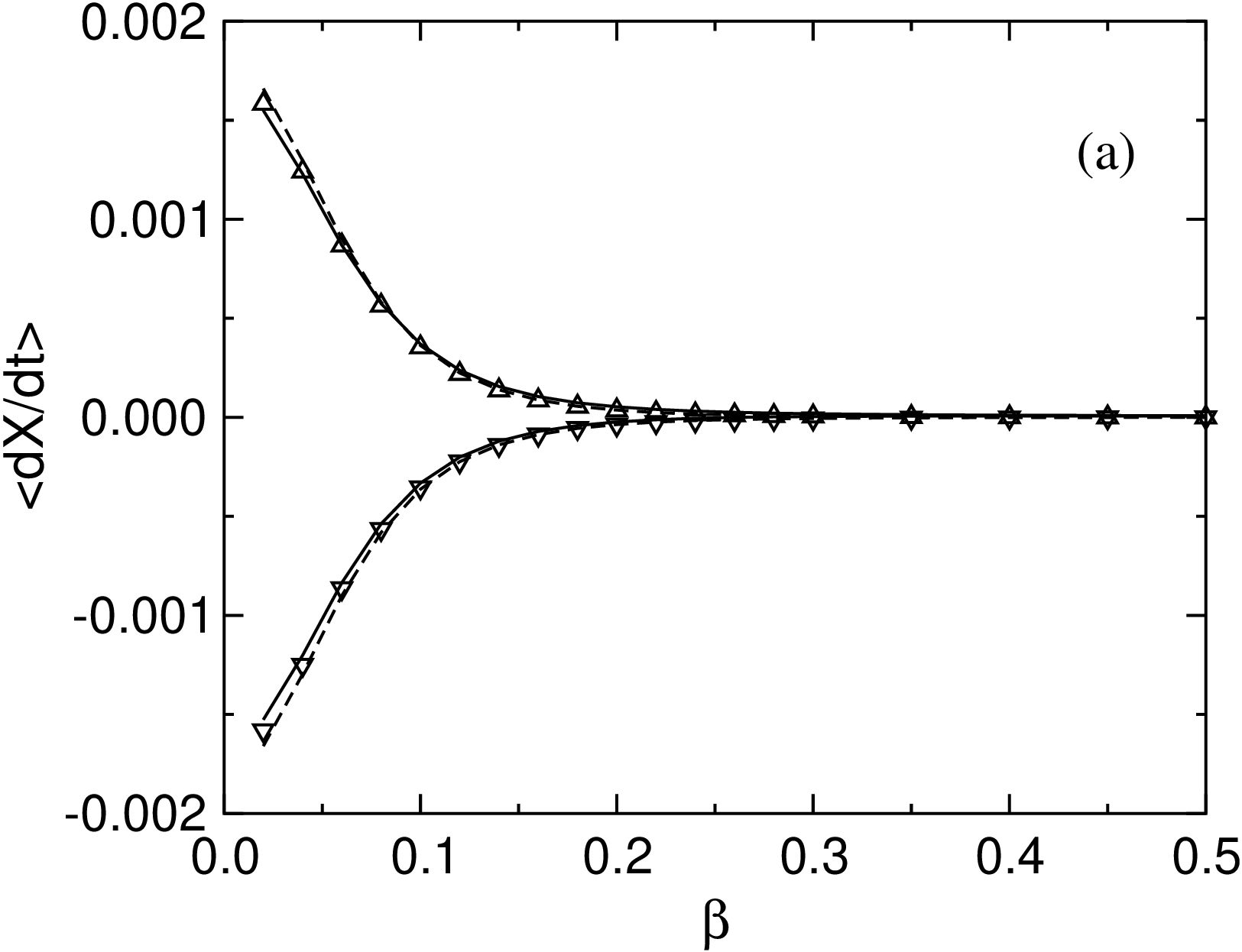}
\includegraphics[width=7.5cm,height=7cm]{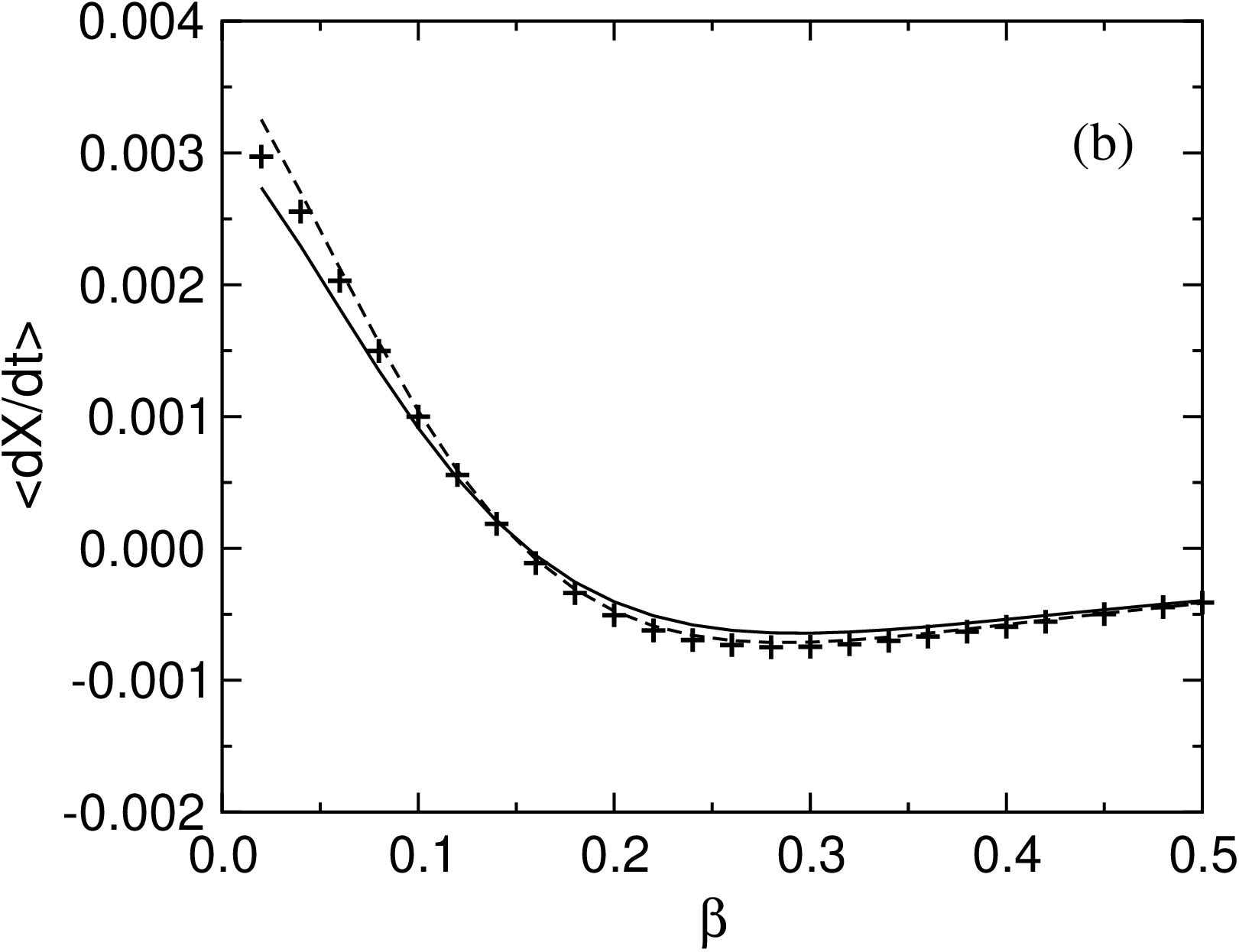}
\end{tabular}
\includegraphics[width=7.5cm,height=7cm]{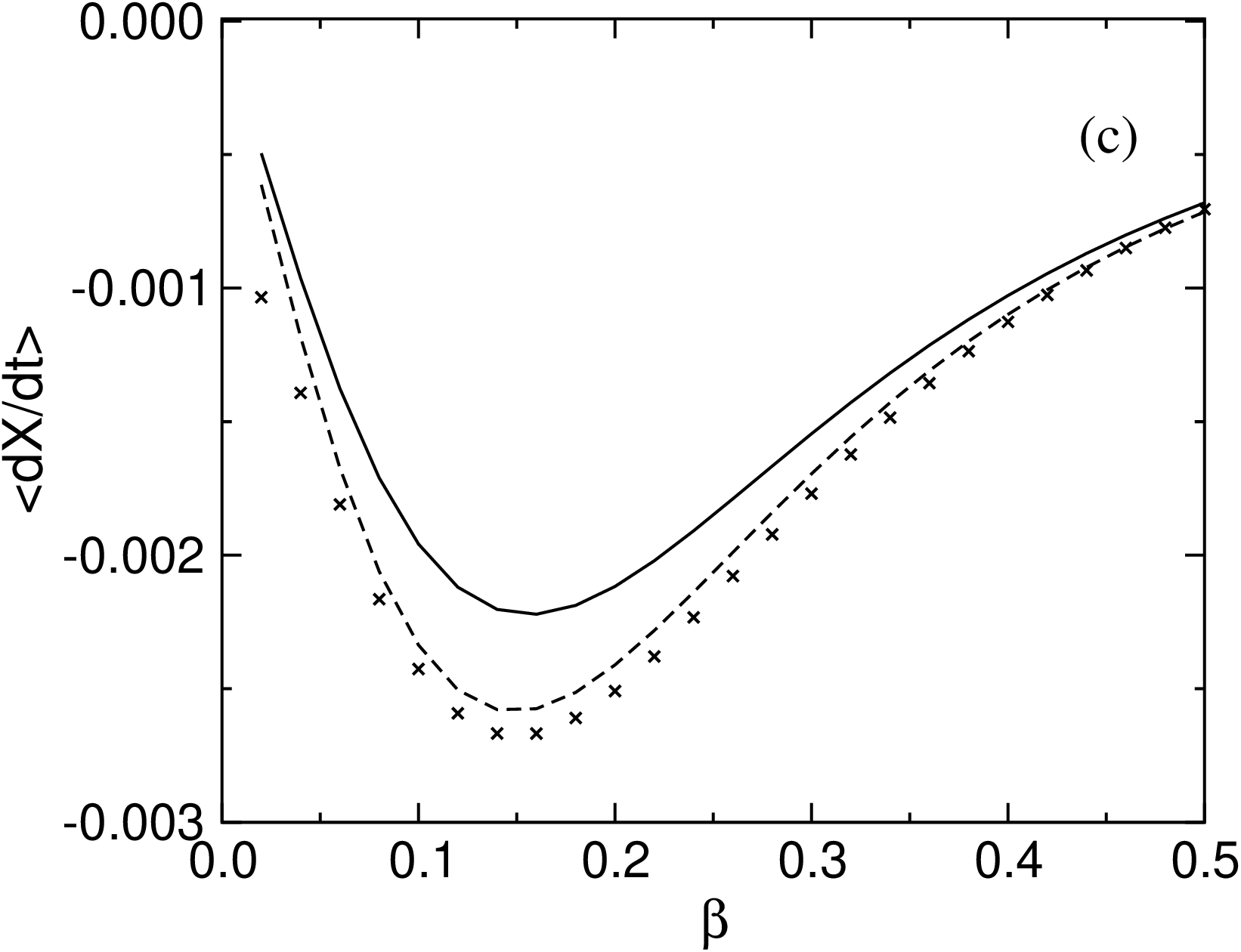}
\end{center}
\caption{Average velocity of a $\phi^4$ kink as a function of the
dissipation for $\beta \ge 0.02$. (a): $\epsilon_{1}=\epsilon_{2}=0.02$, $\delta=0.1$,
$\theta_{1}=\pi/2$. Numerical simulations: $\theta_{2}=\pi$
($\triangledown$); $\theta_{2}=0$ ($\vartriangle$). (b):
$\epsilon_{1}=0.04$, $\epsilon_{2}=0.026$, $\delta=0.25$,
$\theta_{1}=-\pi/2$, $\theta_{2}=-\pi/2+0.8$. Numerical
simulations: (+). (c): $\epsilon_{1}=0.04$, $\epsilon_{2}=0.026$,
$\delta=0.25$, $\theta_{1}=-\pi/2$, $\theta_{2}=-\pi/2$; Numerical
simulations: ($\times$). In all cases the solid lines show results
obtained from the numerical solution of  Eqs.\
(\ref{cc-p})-(\ref{cc-l}) and the dashed lines are the values
obtained from Eq.\ (\ref{2.16}).} \label{phi-beta}
\end{figure}

\subsection{$\phi^4$ model}

After describing in the previous subsection the ratchet dynamics
of kinks in the sG equation, here we focus our attention on the
$\phi^4$ model. This is important in order to ascertain to what
extent does the phenomenology found depend on the unperturbed
equation being integrable, or on the structure of the sG soliton
(which, for instance, does not have internal modes
\cite{modosinternos}, whereas the $\phi^4$ model does have one).
In view of this, we carried out numerical simulations of the
$\phi^4$ model, to compare with the analytical CC results in Sec.\
\ref{CCA-section}. As we will now summarize, the phenomenology of
directed motion of kinks in the $\phi^4$ model is the same as in
the sG model.

To begin with, we can observe in Fig.\ \ref{dftmphi4} that our
first resonance criterion on the synchronization between the
oscillations of $l(t)$ and $f(t)$ does also apply for the $\phi^4$
model. As we can see, for even value of $m$ ($m=2, 4$) there are
resonant frequencies, while in case of $m=3$ the frequencies of
the ac force $\delta$ and $3\, \delta$ never appear in the
spectrum of the width of the kink. Correspondingly, in our
numerical simulations we have found unidirectional motion only in
cases of $m=2, m=4$, in contrast to $m=3$ where an oscillatory
motion of the center of the kink takes place (not shown). The
sinusoidal dependence of the average velocity on the phases is
found as well as in the sG equation. The other feature we focused
on, namely dissipation effects, agrees with the previous picture
as well: In Fig.\ \ref{phi-beta} we observe, as in the sG case,
that changing the damping  we can decrease the average of the
velocity, rectify the motion, or optimize it. In this respect,
another interesting issue is the comparison between the observed
mobility of the kink between the sG and $\phi^4$ models. From Eq.\
(\ref{cc-X}), it can be noticed that, if we consider the first
approximation of $l(t)$, the center of the kink will feel an ac
force with the amplitudes of the two harmonics  modulated by the
factor $\lambda=q/M_{0}$, which in sG is  $\lambda_{sG}=0.785$ and
in $\phi^4$, $\lambda_{\phi4}=2.121$. Therefore, for the same
values of $\epsilon_{i}$ in both systems, the effective amplitude
of the ac force that the center of the kink experiences in
$\phi^4$ is greater than in sG, $\lambda_{sG} \epsilon^{sG}_{i} <
\lambda_{\phi4}\epsilon^{\phi4}_{i}$. Therefore, from the CC
analysis we expect higher mobility for the $\phi^4$ model, which
is in good agreement with the numerical simulations as is shown in
Figs.\  \ref{sG-beta} and \ref{phi-beta}. We thus conclude that
the existence of net motion of kinks (topological excitations) is
a generic phenomenon, both in terms of the criterion for its
appearance and its main characteristics, and not a specific
feature of the sG equation.

\section{Conclusions}

In this paper we have investigated in detail the behavior of
topological excitations when driven by ac forces in homogeneous
systems. We have focused on the nonlinear Klein-Gordon family of
models as a paradigmatic example, and specifically on the sG and
$\phi^4$ equations as examples of integrable and non-integrable
models, respectively. By means of an analytical technique of the
CC class, we have shown that two conditions must be fulfilled in
order to observe unidirectional motion of the kinks: First, the
shift symmetry must be broken, and second the time-reversal
symmetry must be broken as well. In the absence of dissipation,
the first symmetry may be violated by a bi-harmonic force
containing even and odd harmonics of the frequency. We have
devoted a large part of our work to characterize the case in which
the force is composed of the first and second harmonics [$m=2$ in
Eq.\ (\ref{ac})], and we have found that the velocity of the net
motion depends on the relative phase between the two harmonic
forces in a sinusoidal manner. The values of the phase for which
the velocity vanishes (critical phase values) are precisely those
for which, loosely speaking, the breaking of the time-reversal
symmetry by the ac force is compensated by the breaking of the
time-reversal symmetry due to the damping. Numerical simulations
for the two models considered are in very good agreement with the
analytical results, qualitatively in terms of the velocity value,
and quantitatively as far as the critical phase values are
concerned. We have also studied the cases of the third ($m=3$) and
fourth ($m=4$) harmonics, finding, respectively, no motion
(because the shift symmetry is not broken) and motion slower than
in the previous case. Indeed, as we have seen, in general the
efficiency of the driving diminishes when the second harmonic
force is a higher harmonic. In all cases, the arguments in terms
of shift symmetry breaking forces can be equivalently presented as
existence or not of resonant behavior between the frequencies
present in the driving and those in the time evolution of the
width of the kink.

In addition to the breaking of the time-reversal symmetry of the
ac force, we have also studied another form of time-reversal
symmetry violation, namely by introducing dissipation in the
models. In the presence of dissipation, we have observed three
different types of behavior depending on the parameters of the
driving: A monotonically decreasing velocity of the kink with
increasing dissipation, the appearance of an optimum value of the
dissipation with a maximum in the velocity, and even reversals of
the directed motion for some values of the dissipation. In all
three cases, the CC theory describes correctly the phenomena as
observed in the simulation, except when the dissipation is very
small, allowing for a non-negligible excitation of phonons in the
system that breaks down the basic assumption of the CC theory.

The conclusions we have just summarized acquire greater relevance
when compared to different types of experimental results
\cite{schia,JJ2,optical2}. In the absence of dissipation
\cite{schia}, experiments in optical lattices with biharmonic
driving find a sinusoidal dependence on the relative phase similar
to the one reported here. When dissipation is present in Josephson
junctions \cite{JJ2}, the same sinusoidal dependence is again
found, and optimal driving parameters appear. Finally, recent work
on dissipation effects on optical lattices \cite{optical2}
exhibits several of the features we have just discussed, such as
the dependence of the critical phase on the damping, and the
overall sinusoidal dependence. Although the
 motion of cold atoms is essentially based on a one-particle description
 the dissipation in the system is caused by ground state transitions, which
 manifests the presence of an internal dynamics. The role of
 this internal dynamics for the ratchet mechanism in cold
 atoms has been put forward in \cite{Mennerat}.
 In this regard the role of the internal mode in our CC approach
 could shed light on the effect of the interband transitions
 and their disipative effects in the ratchet transport of cold atoms.
We thus see
that our results are very generic, in so far as they apply to a
large family of theoretical models, both integrable and
non-integrable, and they capture the essential ingredients of
experiments in different fields \cite{ricardo}. Therefore, the
main conclusion of our work is that we have correctly identified
the mechanism for the appearance of directed motion due to
zero-mean forces through the breaking of symmetries, and we have
provided a physical interpretation of this mechanism in terms of
the coupling between excitation modes of the system. We hope that
this work encourages further experimental work to check the
remaining features discussed.


\begin{acknowledgments}
This work has been supported by the Ministerio de Educaci\'on y
Ciencia (MEC, Spain) and DAAD (Germany) through ``Acciones
Integradas Hispano-Alemanas'' HA2004-0034---D/04/39957, by MEC
grants FIS2005-973 (NRQ), BFM2003-07749-C05-01, FIS2004-01001 and
NAN2004-09087-C03-03 (AS), and by the Junta de Andaluc\'{\i}a
under the project FQM-0207.
\end{acknowledgments}

\end{document}